\def\DIRvalue{Tachikawa}
\def\IDvalue{TA}
\def\titlevalue{A brief review of the 2d/4d correspondences}
\def\authorvalue{Yuji Tachikawa}
\def\shortauthorvalue{\authorvalue}
\def\addressvalue{ Kavli Institute for the Physics and Mathematics of the Universe, \\
 University of Tokyo,  Kashiwa, Chiba 277-8583, Japan\\
  \tt yuji.tachikawa@ipmu.jp}
\def\abstractvalue{ An elementary introduction to the 2d/4d correspondences is given. 
After quickly reviewing the 2d $q$-deformed Yang-Mills theory and the Liouville theory,
we will introduce 4d theories obtained by coupling trifundamentals to SU(2) gauge fields.
We will then see concretely that the supersymmetric partition function of these theories on $S^3\times S^1$ and on $S^4$ is given respectively by the $q$-deformed Yang-Mills theory and the Liouville theory.
After giving a short discussion on how this correspondence may be understood from the viewpoint of the 6d $\mathcal{N}=(2,0)$ theory, we conclude the review by enumerating future directions. 
Most of the technical points will be referred to more detailed review articles. 
}
\def\preprintvalue{IPMU-16-0048}

\ifx\ifLONG\undefined
\documentclass[12pt]{article}
\newcommand{\chapterauthor}[1]{
\begin{center}
{\bf \normalsize  #1}
\end{center}
}

\newcommand{\chapteraddress}[1]{
\begin{center}
{ \small \it \addressvalue}
\end{center}
}

\newcommand{\chapterabstract}[1]{
\vspace{\baselineskip}
\begin{center}
\textbf{\small Abstract}
\end{center}
#1}

\newcommand{\chapterheader}{

\chapter[\titlevalue{}  (by \shortauthorvalue)]{\titlevalue}
\label{Chapter\IDvalue}
\chapterauthor{\authorvalue}
\chapteraddress{\addressvalue}
\chapterabstract{\abstractvalue}
\tightmtctrue
\minitoc
}

\newcommand{\documentheader}{
\begin{flushright} \small
  \preprintvalue
 \end{flushright}

\begin{center}
{\bf \Large \titlevalue}
\end{center}

\chapterauthor{\authorvalue}
\chapteraddress{\addressvalue}
\chapterabstract{\abstractvalue}

\medskip

This is a contribution to the review volume ``Localization techniques
in quantum field theories'' (eds. V.~Pestun and M.~Zabzine) which
contains 17 Chapters available at \cite{ContributionSummary}

\tableofcontents
}

\newcommand{\ifvolume}[2]{\ifx\ifLONG\undefined#2\else#1\fi}

\newcommand{\documentfinish}{
\ifx\ifLONG\undefined
\bibliographystyle{bibreview} 
\bibliography{\IDvalue,review}  
\end{document}
\else
\addcontentsline{toc}{section}{References}
\providecommand{\href}[2]{#2}\begingroup\raggedright\begin{thebibliography}{10}

\bibitem{ContributionSummary}
V.~Pestun and M.~Zabzine, eds., {\em Localization techniques in quantum field
  theory}, vol.~xx.
\newblock Journal of Physics A, 2016.
\newblock \href{http://arxiv.org/abs/1608.02952}{{\tt 1608.02952}}.
\newblock \url{https://arxiv.org/src/1608.02952/anc/LocQFT.pdf},
  \url{http://pestun.ihes.fr/pages/LocalizationReview/LocQFT.pdf}.

\bibitem{TAAlday:2009aq}
L.~F. Alday, D.~Gaiotto, and Y.~Tachikawa, ``{Liouville Correlation Functions
  from Four-Dimensional Gauge Theories},''
  \href{http://dx.doi.org/10.1007/s11005-010-0369-5}{{\em Lett. Math. Phys.}
  {\bf 91} (2010)  167--197},
\href{http://arxiv.org/abs/0906.3219}{{\tt arXiv:0906.3219 [hep-th]}}.
\bibitem{TAGadde:2011ik}
A.~Gadde, L.~Rastelli, S.~S. Razamat, and W.~Yan, ``{The 4D Superconformal
  Index from $q$-Deformed 2D Yang- Mills},''
  \href{http://dx.doi.org/10.1103/PhysRevLett.106.241602}{{\em Phys. Rev.
  Lett.} {\bf 106} (2011)  241602},
\href{http://arxiv.org/abs/1104.3850}{{\tt arXiv:1104.3850 [hep-th]}}.
\bibitem{TAGaiotto:2009we}
D.~Gaiotto, ``{${\mathcal{N}}=2$ Dualities},''
  \href{http://dx.doi.org/10.1007/JHEP08(2012)034}{{\em JHEP} {\bf 1208} (2012)
   034},
\href{http://arxiv.org/abs/0904.2715}{{\tt arXiv:0904.2715 [hep-th]}}.
\bibitem{TANekrasov:2002qd}
N.~A. Nekrasov, ``{Seiberg-Witten Prepotential from Instanton Counting},'' {\em
  Adv. Theor. Math. Phys.} {\bf 7} (2004)  831--864,
\href{http://arxiv.org/abs/hep-th/0206161}{{\tt arXiv:hep-th/0206161}}.
\bibitem{TAPestun:2007rz}
V.~Pestun, ``{Localization of Gauge Theory on a Four-Sphere and Supersymmetric
  Wilson Loops},'' \href{http://dx.doi.org/10.1007/s00220-012-1485-0}{{\em
  Commun.Math.Phys.} {\bf 313} (2012)  71--129},
\href{http://arxiv.org/abs/0712.2824}{{\tt arXiv:0712.2824 [hep-th]}}.
\bibitem{TARomelsberger:2005eg}
C.~Romelsberger, ``{Counting Chiral Primaries in ${\mathcal{N}}\!=1$, $d=4$
  Superconformal Field Theories},''
  \href{http://dx.doi.org/10.1016/j.nuclphysb.2006.03.037}{{\em Nucl.Phys.}
  {\bf B747} (2006)  329--353},
\href{http://arxiv.org/abs/hep-th/0510060}{{\tt arXiv:hep-th/0510060
  [hep-th]}}.
\bibitem{TAKinney:2005ej}
J.~Kinney, J.~M. Maldacena, S.~Minwalla, and S.~Raju, ``{An Index for 4
  Dimensional Super Conformal Theories},''
  \href{http://dx.doi.org/10.1007/s00220-007-0258-7}{{\em Commun.Math.Phys.}
  {\bf 275} (2007)  209--254},
\href{http://arxiv.org/abs/hep-th/0510251}{{\tt arXiv:hep-th/0510251
  [hep-th]}}.
\bibitem{TAGadde:2011uv}
A.~Gadde, L.~Rastelli, S.~S. Razamat, and W.~Yan, ``{Gauge Theories and
  Macdonald Polynomials},''
  \href{http://dx.doi.org/10.1007/s00220-012-1607-8}{{\em Commun.Math.Phys.}
  {\bf 319} (2013)  147--193},
\href{http://arxiv.org/abs/1110.3740}{{\tt arXiv:1110.3740 [hep-th]}}.
\bibitem{TACordes:1994fc}
S.~Cordes, G.~W. Moore, and S.~Ramgoolam, ``{Lectures on 2-D Yang-Mills Theory,
  Equivariant Cohomology and Topological Field Theories},''
  \href{http://dx.doi.org/10.1016/0920-5632(95)00434-B}{{\em
  Nucl.Phys.Proc.Suppl.} {\bf 41} (1995)  184--244},
\href{http://arxiv.org/abs/hep-th/9411210}{{\tt arXiv:hep-th/9411210
  [hep-th]}}.
\bibitem{TABuffenoir:1994fh}
E.~Buffenoir and P.~Roche, ``{Two-Dimensional Lattice Gauge Theory Based on a
  Quantum Group},'' \href{http://dx.doi.org/10.1007/BF02099153}{{\em
  Commun.Math.Phys.} {\bf 170} (1995)  669--698},
\href{http://arxiv.org/abs/hep-th/9405126}{{\tt arXiv:hep-th/9405126
  [hep-th]}}.
\bibitem{TAAlekseev:1994au}
A.~Y. Alekseev, H.~Grosse, and V.~Schomerus, ``{Combinatorial Quantization of
  the Hamiltonian Chern-Simons Theory. 2.},''
  \href{http://dx.doi.org/10.1007/BF02101528}{{\em Commun.Math.Phys.} {\bf 174}
  (1995)  561--604},
\href{http://arxiv.org/abs/hep-th/9408097}{{\tt arXiv:hep-th/9408097
  [hep-th]}}.
\bibitem{TAAganagic:2004js}
M.~Aganagic, H.~Ooguri, N.~Saulina, and C.~Vafa, ``{Black Holes, $q$-Deformed
  2D Yang-Mills, and Non-Perturbative Topological Strings},''
  \href{http://dx.doi.org/10.1016/j.nuclphysb.2005.02.035}{{\em Nucl.Phys.}
  {\bf B715} (2005)  304--348},
\href{http://arxiv.org/abs/hep-th/0411280}{{\tt arXiv:hep-th/0411280
  [hep-th]}}.
\bibitem{TASeiberg:1990eb}
N.~Seiberg, ``{Notes on Quantum Liouville Theory and Quantum Gravity},''
\href{http://dx.doi.org/10.1143/PTPS.102.319}{{\em Prog. Theor. Phys. Suppl.}
  {\bf 102} (1990)  319--349}.
\bibitem{TANakayama:2004vk}
Y.~Nakayama, ``{Liouville Field Theory: a Decade After the Revolution},''
  \href{http://dx.doi.org/10.1142/S0217751X04019500}{{\em Int. J. Mod. Phys.}
  {\bf A19} (2004)  2771--2930},
\href{http://arxiv.org/abs/hep-th/0402009}{{\tt arXiv:hep-th/0402009}}.
\bibitem{TATeschner:2001rv}
J.~Teschner, ``{Liouville theory revisited},'' {\em Class. Quant. Grav.} {\bf
  18} (2001)  R153--R222,
\href{http://arxiv.org/abs/hep-th/0104158}{{\tt arXiv:hep-th/0104158}}.
\bibitem{TADorn:1994xn}
H.~Dorn and H.~J. Otto, ``{Two and Three Point Functions in Liouville
  Theory},'' \href{http://dx.doi.org/10.1016/0550-3213(94)00352-1}{{\em Nucl.
  Phys.} {\bf B429} (1994)  375--388},
\href{http://arxiv.org/abs/hep-th/9403141}{{\tt arXiv:hep-th/9403141}}.
\bibitem{TAZamolodchikov:1995aa}
A.~B. Zamolodchikov and A.~B. Zamolodchikov, ``{Structure Constants and
  Conformal Bootstrap in Liouville Field Theory},''
  \href{http://dx.doi.org/10.1016/0550-3213(96)00351-3}{{\em Nucl. Phys.} {\bf
  B477} (1996)  577--605},
\href{http://arxiv.org/abs/hep-th/9506136}{{\tt arXiv:hep-th/9506136}}.
\bibitem{TAPonsot:1999uf}
B.~Ponsot and J.~Teschner, ``{Liouville Bootstrap via Harmonic Analysis on a
  Noncompact Quantum Group},''
\href{http://arxiv.org/abs/hep-th/9911110}{{\tt arXiv:hep-th/9911110}}.
\bibitem{TATachikawa:2013kta}
Y.~Tachikawa, ``{$\mathcal{N}{=}2$ supersymmetric dynamics for pedestrians},''
  \href{http://dx.doi.org/10.1007/978-3-319-08822-8}{{\em Lect. Notes Phys.}
  {\bf 890} (2013)  194}, \href{http://arxiv.org/abs/1312.2684}{{\tt
  arXiv:1312.2684 [hep-th]}}.
\url{http://inspirehep.net/record/1268680/files/arXiv:1312.2684.pdf}.
\bibitem{TAFestuccia:2011ws}
G.~Festuccia and N.~Seiberg, ``{Rigid Supersymmetric Theories in Curved
  Superspace},'' \href{http://dx.doi.org/10.1007/JHEP06(2011)114}{{\em JHEP}
  {\bf 1106} (2011)  114},
\href{http://arxiv.org/abs/1105.0689}{{\tt arXiv:1105.0689 [hep-th]}}.
\bibitem{ContributionDU}
T.~Dumitrescu, ``An Introduction to Supersymmetric Field Theories in Curved
  Space,'' {\em Journal of Physics A} {\bf xx} (2016)  000,
  \href{http://arxiv.org/abs/1608.02957}{{\tt 1608.02957}}.

\bibitem{ContributionRR}
L.~Rastelli and S.~Razamat, ``The supersymmetric index in four dimensions,''
  {\em Journal of Physics A} {\bf xx} (2016)  000,
  \href{http://arxiv.org/abs/1608.02965}{{\tt 1608.02965}}.

\bibitem{ContributionHO}
K.~Hosomichi, ``$\mathcal{N}=2$ SUSY gauge theories on $S^4$,'' {\em Journal of
  Physics A} {\bf xx} (2016)  000, \href{http://arxiv.org/abs/1608.02962}{{\tt
  1608.02962}}.

\bibitem{TANawata:2011un}
S.~Nawata, ``{Localization of ${\mathcal{N}}\!=4$ Superconformal Field Theory
  on $S^1\times S^3$ and Index},''
  \href{http://dx.doi.org/10.1007/JHEP11(2011)144}{{\em JHEP} {\bf 1111} (2011)
   144},
\href{http://arxiv.org/abs/1104.4470}{{\tt arXiv:1104.4470 [hep-th]}}.
\bibitem{TAHama:2012bg}
N.~Hama and K.~Hosomichi, ``{Seiberg-Witten Theories on Ellipsoids},''
  \href{http://dx.doi.org/10.1007/JHEP09(2012)033,
  10.1007/JHEP10(2012)051}{{\em JHEP} {\bf 1209} (2012)  033},
\href{http://arxiv.org/abs/1206.6359}{{\tt arXiv:1206.6359 [hep-th]}}.
\bibitem{TAPestunLocalizationReview}
V.~Pestun, \href{http://dx.doi.org/10.1007/978-3-319-18769-3_6}{``{Localization
  for $\mathcal {N}=2$ Supersymmetric Gauge Theories in Four Dimensions},''} in
  {\em New Dualities of Supersymmetric Gauge Theories}, J.~Teschner, ed.,
  pp.~159--194.
\newblock Springer, 2016.
\newblock
\href{http://arxiv.org/abs/1412.7134}{{\tt arXiv:1412.7134 [hep-th]}}.
\newblock
\bibitem{TATachikawaInstantonReview}
Y.~Tachikawa, \href{http://dx.doi.org/10.1007/978-3-319-18769-3_4}{``{A review
  on instanton counting and W-algebras},''} in {\em New Dualities of
  Supersymmetric Gauge Theories}, J.~Teschner, ed., pp.~79--120.
\newblock Springer, 2016.
\newblock
\href{http://arxiv.org/abs/1412.7121}{{\tt arXiv:1412.7121 [hep-th]}}.
\newblock
\bibitem{TAOkuda:2010ke}
T.~Okuda and V.~Pestun, ``{On the Instantons and the Hypermultiplet Mass of
  ${\mathcal{N}}\!=2$* Super Yang-Mills on $S^4$},''
  \href{http://dx.doi.org/10.1007/JHEP03(2012)017}{{\em JHEP} {\bf 1203} (2012)
   017},
\href{http://arxiv.org/abs/1004.1222}{{\tt arXiv:1004.1222 [hep-th]}}.
\bibitem{TAErickson:2000af}
J.~Erickson, G.~Semenoff, and K.~Zarembo, ``{Wilson Loops in
  ${\mathcal{N}}\!=4$ Supersymmetric Yang-Mills Theory},''
  \href{http://dx.doi.org/10.1016/S0550-3213(00)00300-X}{{\em Nucl.Phys.} {\bf
  B582} (2000)  155--175},
\href{http://arxiv.org/abs/hep-th/0003055}{{\tt arXiv:hep-th/0003055
  [hep-th]}}.
\bibitem{TADrukker:2000rr}
N.~Drukker and D.~J. Gross, ``{An Exact Prediction of ${\mathcal{N}}\!=4$ SUSYM
  Theory for String Theory},'' \href{http://dx.doi.org/10.1063/1.1372177}{{\em
  J.Math.Phys.} {\bf 42} (2001)  2896--2914},
\href{http://arxiv.org/abs/hep-th/0010274}{{\tt arXiv:hep-th/0010274
  [hep-th]}}.
\bibitem{TAGerchkovitz:2014gta}
E.~Gerchkovitz, J.~Gomis, and Z.~Komargodski, ``{Sphere Partition Functions and
  the Zamolodchikov Metric},''
  \href{http://dx.doi.org/10.1007/JHEP11(2014)001}{{\em JHEP} {\bf 1411} (2014)
   001},
\href{http://arxiv.org/abs/1405.7271v2}{{\tt arXiv:1405.7271v2 [hep-th]}}.
\bibitem{TAGomis:2014woa}
J.~Gomis and N.~Ishtiaque, ``{K\"ahler Potential and Ambiguities in 4D
  ${\mathcal{N}}\!=2$ SCFTs},''
\href{http://arxiv.org/abs/1409.5325}{{\tt arXiv:1409.5325 [hep-th]}}.
\bibitem{TAGomis:2015yaa}
J.~Gomis, P.-S. Hsin, Z.~Komargodski, A.~Schwimmer, N.~Seiberg, and S.~Theisen,
  ``{Anomalies, Conformal Manifolds, and Spheres},''
  \href{http://dx.doi.org/10.1007/JHEP03(2016)022}{{\em JHEP} {\bf 03} (2016)
  022},
\href{http://arxiv.org/abs/1509.08511}{{\tt arXiv:1509.08511 [hep-th]}}.
\bibitem{ContributionKL}
S.~Kim and K.~Lee, ``Indices for 6 dimensional superconformal field theories,''
  {\em Journal of Physics A} {\bf xx} (2016)  000,
  \href{http://arxiv.org/abs/1608.02969}{{\tt 1608.02969}}.

\bibitem{TAFukuda:2012jr}
Y.~Fukuda, T.~Kawano, and N.~Matsumiya, ``{5D SYM and 2D $q$-Deformed YM},''
\href{http://arxiv.org/abs/1210.2855}{{\tt arXiv:1210.2855 [hep-th]}}.
\bibitem{TACordovaJafferis}
C.~Cordova and D.~L. Jafferis, ``{Toda Theory From Six Dimensions},''
\href{http://arxiv.org/abs/1605.03997}{{\tt arXiv:1605.03997 [hep-th]}}.
\bibitem{TAHollands:2011zc}
L.~Hollands, C.~A. Keller, and J.~Song, ``{Towards a 4D/2D Correspondence for
  Sicilian Quivers},'' \href{http://dx.doi.org/10.1007/JHEP10(2011)100}{{\em
  JHEP} {\bf 10} (2011)  100},
\href{http://arxiv.org/abs/1107.0973}{{\tt arXiv:1107.0973 [hep-th]}}.
\bibitem{TARazamat:2013qfa}
S.~S. Razamat, ``{On the $\mathcal{N} = 2$ Superconformal Index and
  Eigenfunctions of the Elliptic RS Model},''
  \href{http://dx.doi.org/10.1007/s11005-014-0682-5}{{\em Lett.Math.Phys.} {\bf
  104} (2014)  673--690},
\href{http://arxiv.org/abs/1309.0278}{{\tt arXiv:1309.0278 [hep-th]}}.
\bibitem{TAGaiotto:2012sf}
D.~Gaiotto and J.~Teschner, ``{Irregular Singularities in Liouville Theory and
  Argyres-Douglas Type Gauge Theories, I},''
\href{http://arxiv.org/abs/1203.1052}{{\tt arXiv:1203.1052 [hep-th]}}.
\bibitem{TAGomis:2014eya}
J.~Gomis and B.~Le~Floch, ``{M2-brane surface operators and gauge theory
  dualities in Toda},''
\href{http://arxiv.org/abs/1407.1852}{{\tt arXiv:1407.1852 [hep-th]}}.
\bibitem{TABuican:2015ina}
M.~Buican and T.~Nishinaka, ``{On the superconformal index of Argyres--Douglas
  theories},'' \href{http://dx.doi.org/10.1088/1751-8113/49/1/015401}{{\em J.
  Phys.} {\bf A49} (2016) no.~1, 015401},
\href{http://arxiv.org/abs/1505.05884}{{\tt arXiv:1505.05884 [hep-th]}}.
\bibitem{TACordova:2015nma}
C.~Cordova and S.-H. Shao, ``{Schur Indices, BPS Particles, and Argyres-Douglas
  Theories},'' \href{http://dx.doi.org/10.1007/JHEP01(2016)040}{{\em JHEP} {\bf
  01} (2016)  040},
\href{http://arxiv.org/abs/1506.00265}{{\tt arXiv:1506.00265 [hep-th]}}.
\bibitem{TAAlday:2013kda}
L.~F. Alday, M.~Bullimore, M.~Fluder, and L.~Hollands, ``{Surface Defects, the
  Superconformal Index and $q$-Deformed Yang-Mills},''
\href{http://arxiv.org/abs/1303.4460}{{\tt arXiv:1303.4460 [hep-th]}}.
\bibitem{TAAlday:2009fs}
L.~F. Alday, D.~Gaiotto, S.~Gukov, Y.~Tachikawa, and H.~Verlinde, ``{Loop and
  Surface Operators in ${\mathcal{N}}\!=2$ Gauge Theory and Liouville Modular
  Geometry},'' \href{http://dx.doi.org/10.1007/JHEP01(2010)113}{{\em JHEP} {\bf
  1001} (2010)  113},
\href{http://arxiv.org/abs/0909.0945}{{\tt arXiv:0909.0945 [hep-th]}}.
\bibitem{TADrukker:2009id}
N.~Drukker, J.~Gomis, T.~Okuda, and J.~Teschner, ``{Gauge Theory Loop Operators
  and Liouville Theory},''
  \href{http://dx.doi.org/10.1007/JHEP02(2010)057}{{\em JHEP} {\bf 02} (2010)
  057},
\href{http://arxiv.org/abs/0909.1105}{{\tt arXiv:0909.1105 [hep-th]}}.
\bibitem{TANawataToAppear}
S.~Nawata, ``{Givental J-functions, Quantum integrable systems, AGT relation
  with surface operator},''
\href{http://arxiv.org/abs/1408.4132}{{\tt arXiv:1408.4132 [hep-th]}}.
\bibitem{TAAlday:2010vg}
L.~F. Alday and Y.~Tachikawa, ``{Affine $SL(2)$ Conformal Blocks from 4D Gauge
  Theories},'' \href{http://dx.doi.org/10.1007/s11005-010-0422-4}{{\em
  Lett.Math.Phys.} {\bf 94} (2010)  87--114},
\href{http://arxiv.org/abs/1005.4469}{{\tt arXiv:1005.4469 [hep-th]}}.
\bibitem{TAXie:2012hs}
D.~Xie, ``{General Argyres-Douglas Theory},''
  \href{http://dx.doi.org/10.1007/JHEP01(2013)100}{{\em JHEP} {\bf 1301} (2013)
   100},
\href{http://arxiv.org/abs/1204.2270}{{\tt arXiv:1204.2270 [hep-th]}}.
\bibitem{TAChacaltana:2012zy}
O.~Chacaltana, J.~Distler, and Y.~Tachikawa, ``{Nilpotent Orbits and
  Codimension-Two Defects of 6D N=(2,0) Theories},''
  \href{http://dx.doi.org/10.1142/S0217751X1340006X}{{\em Int.J.Mod.Phys.} {\bf
  A28} (2013)  1340006},
\href{http://arxiv.org/abs/1203.2930}{{\tt arXiv:1203.2930 [hep-th]}}.
\bibitem{TASchiffmannVasserot}
O.~Schiffmann and E.~Vasserot, ``{Cherednik algebras, W algebras and the
  equivariant cohomology of the moduli space of instantons on
  $\mathbb{A}^2$},'' \href{http://dx.doi.org/10.1007/s10240-013-0052-3}{{\em
  Publications math\'ematiques de l'IH\'ES} {\bf 118} (2013) no.~1, 213--342},
  \href{http://arxiv.org/abs/1202.2756}{{\tt arXiv:1202.2756 [math.QA]}}.

\bibitem{TAMaulik:2012wi}
D.~Maulik and A.~Okounkov, ``{Quantum Groups and Quantum Cohomology},''
\href{http://arxiv.org/abs/1211.1287}{{\tt arXiv:1211.1287 [math.AG]}}.
\bibitem{TABraverman:2014xca}
A.~Braverman, M.~Finkelberg, and H.~Nakajima, ``{Instanton Moduli Spaces and
  $\mathcal{W}$-Algebras},''
\href{http://arxiv.org/abs/1406.2381}{{\tt arXiv:1406.2381 [math.QA]}}.
\bibitem{TANekrasovLisbon}
N.~{Nekrasov},
  \href{http://dx.doi.org/10.1142/9789812704016_0066}{``{Localizing gauge
  theories},''} in {\em XIVth International Congress on Mathematical Physics},
  J.-C. {Zambrini}, ed., pp.~645--654.
\newblock Mar., 2006.
\newblock \url{http://www.ihes.fr/~nikita/IMAGES/Lisbon.ps}.

\bibitem{TABershtein:2015xfa}
M.~Bershtein, G.~Bonelli, M.~Ronzani, and A.~Tanzini, ``{Exact results for
  ${\cal N}=2$ supersymmetric gauge theories on compact toric manifolds and
  equivariant Donaldson invariants},''
\href{http://arxiv.org/abs/1509.00267}{{\tt arXiv:1509.00267 [hep-th]}}.
\bibitem{TAFucito:2004ry}
F.~Fucito, J.~F. Morales, and R.~Poghossian, ``{Multi instanton calculus on ALE
  spaces},'' \href{http://dx.doi.org/10.1016/j.nuclphysb.2004.09.014}{{\em
  Nucl. Phys.} {\bf B703} (2004)  518--536},
\href{http://arxiv.org/abs/hep-th/0406243}{{\tt arXiv:hep-th/0406243}}.
\bibitem{TAFucito:2006kn}
F.~Fucito, J.~F. Morales, and R.~Poghossian, ``{Instanton on toric
  singularities and black hole countings},''
  \href{http://dx.doi.org/10.1088/1126-6708/2006/12/073}{{\em JHEP} {\bf 12}
  (2006)  073},
\href{http://arxiv.org/abs/hep-th/0610154}{{\tt arXiv:hep-th/0610154
  [hep-th]}}.
\bibitem{TAGriguolo:2006kp}
L.~Griguolo, D.~Seminara, R.~J. Szabo, and A.~Tanzini, ``{Black holes,
  instanton counting on toric singularities and q-deformed two-dimensional
  Yang-Mills theory},''
  \href{http://dx.doi.org/10.1016/j.nuclphysb.2007.02.030}{{\em Nucl. Phys.}
  {\bf B772} (2007)  1--24},
\href{http://arxiv.org/abs/hep-th/0610155}{{\tt arXiv:hep-th/0610155
  [hep-th]}}.
\bibitem{TADijkgraaf:2007fe}
R.~Dijkgraaf and P.~Sulkowski, ``{Instantons on ALE spaces and orbifold
  partitions},'' \href{http://dx.doi.org/10.1088/1126-6708/2008/03/013}{{\em
  JHEP} {\bf 03} (2008)  013},
\href{http://arxiv.org/abs/0712.1427}{{\tt arXiv:0712.1427 [hep-th]}}.
\bibitem{TABelavin:2011pp}
V.~Belavin and B.~Feigin, ``{Super Liouville conformal blocks from N=2 SU(2)
  quiver gauge theories},''
  \href{http://dx.doi.org/10.1007/JHEP07(2011)079}{{\em JHEP} {\bf 07} (2011)
  079},
\href{http://arxiv.org/abs/1105.5800}{{\tt arXiv:1105.5800 [hep-th]}}.
\bibitem{TABonelli:2011jx}
G.~Bonelli, K.~Maruyoshi, and A.~Tanzini, ``{Instantons on ALE spaces and Super
  Liouville Conformal Field Theories},''
\href{http://arxiv.org/abs/1106.2505}{{\tt arXiv:1106.2505 [hep-th]}}.
\bibitem{TABelavin:2011tb}
A.~Belavin, V.~Belavin, and M.~Bershtein, ``{Instantons and 2d Superconformal
  field theory},'' \href{http://dx.doi.org/10.1007/JHEP09(2011)117}{{\em JHEP}
  {\bf 1109} (2011)  117},
\href{http://arxiv.org/abs/1106.4001}{{\tt arXiv:1106.4001 [hep-th]}}.
\bibitem{TABonelli:2011kv}
G.~Bonelli, K.~Maruyoshi, and A.~Tanzini, ``{Gauge Theories on ALE Space and
  Super Liouville Correlation Functions},''
\href{http://arxiv.org/abs/1107.4609}{{\tt arXiv:1107.4609 [hep-th]}}.
\bibitem{TAWyllard:2011mn}
N.~Wyllard, ``{Coset conformal blocks and N=2 gauge theories},''
\href{http://arxiv.org/abs/1109.4264}{{\tt arXiv:1109.4264 [hep-th]}}.
\bibitem{TAIto:2011mw}
Y.~Ito, ``{Ramond sector of super Liouville theory from instantons on an ALE
  space},'' \href{http://dx.doi.org/10.1016/j.nuclphysb.2012.04.001}{{\em
  Nucl.Phys.} {\bf B861} (2012)  387--402},
\href{http://arxiv.org/abs/1110.2176}{{\tt arXiv:1110.2176 [hep-th]}}.
\bibitem{TAAlfimov:2011ju}
M.~Alfimov and G.~Tarnopolsky, ``{Parafermionic Liouville field theory and
  instantons on ALE spaces},''
  \href{http://dx.doi.org/10.1007/JHEP02(2012)036}{{\em JHEP} {\bf 1202} (2012)
   036},
\href{http://arxiv.org/abs/1110.5628}{{\tt arXiv:1110.5628 [hep-th]}}.
\bibitem{TABelavin:2011sw}
A.~Belavin, M.~Bershtein, B.~Feigin, A.~Litvinov, and G.~Tarnopolsky,
  ``{Instanton moduli spaces and bases in coset conformal field theory},''
\href{http://arxiv.org/abs/1111.2803}{{\tt arXiv:1111.2803 [hep-th]}}.
\bibitem{TABonelli:2012ny}
G.~Bonelli, K.~Maruyoshi, A.~Tanzini, and F.~Yagi, ``{N=2 gauge theories on
  toric singularities, blow-up formulae and W-algebrae},''
\href{http://arxiv.org/abs/1208.0790}{{\tt arXiv:1208.0790 [hep-th]}}.
\bibitem{TABelavin:2012eg}
A.~Belavin, M.~Bershtein, and G.~Tarnopolsky, ``{Bases in coset conformal field
  theory from AGT correspondence and Macdonald polynomials at the roots of
  unity},''
\href{http://arxiv.org/abs/1211.2788}{{\tt arXiv:1211.2788 [hep-th]}}.
\bibitem{TAIto:2013kpa}
Y.~Ito, K.~Maruyoshi, and T.~Okuda, ``{Scheme dependence of instanton counting
  in ALE spaces},''
\href{http://arxiv.org/abs/1303.5765}{{\tt arXiv:1303.5765 [hep-th]}}.
\bibitem{TAAlfimov:2013cqa}
M.~N. Alfimov, A.~A. Belavin, and G.~M. Tarnopolsky, ``{Coset conformal field
  theory and instanton counting on $C^{2}/Z_{p}$},''
  \href{http://dx.doi.org/10.1007/JHEP08(2013)134}{{\em JHEP} {\bf 08} (2013)
  134},
\href{http://arxiv.org/abs/1306.3938}{{\tt arXiv:1306.3938 [hep-th]}}.
\bibitem{TAItoyama:2013mca}
H.~Itoyama, T.~Oota, and R.~Yoshioka, ``{2d-4d Connection between
  $q$-Virasoro/W Block at Root of Unity Limit and Instanton Partition Function
  on ALE Space},''
  \href{http://dx.doi.org/10.1016/j.nuclphysb.2013.10.012}{{\em Nucl. Phys.}
  {\bf B877} (2013)  506--537},
\href{http://arxiv.org/abs/1308.2068}{{\tt arXiv:1308.2068 [hep-th]}}.
\bibitem{TABruzzo:2013daa}
U.~Bruzzo, M.~Pedrini, F.~Sala, and R.~J. Szabo, ``{Framed sheaves on root
  stacks and supersymmetric gauge theories on ALE spaces},'' {\em Adv. Math.}
  {\bf 288} (2016)  1175--1308,
\href{http://arxiv.org/abs/1312.5554}{{\tt arXiv:1312.5554 [math.AG]}}.
\bibitem{TAPedrini:2014yoa}
M.~Pedrini, F.~Sala, and R.~J. Szabo, ``{AGT relations for abelian quiver gauge
  theories on ALE spaces},''
  \href{http://dx.doi.org/10.1016/j.geomphys.2016.01.004}{{\em J. Geom. Phys.}
  {\bf 103} (2016)  43--89},
\href{http://arxiv.org/abs/1405.6992}{{\tt arXiv:1405.6992 [math.RT]}}.
\bibitem{TASpodyneiko:2014qsa}
L.~Spodyneiko, ``{AGT correspondence: Ding-Iohara algebra at roots of unity and
  Lepowsky-Wilson construction},''
  \href{http://dx.doi.org/10.1088/1751-8113/48/27/275404}{{\em J. Phys.} {\bf
  A48} (2015)  275404},
\href{http://arxiv.org/abs/1409.3465}{{\tt arXiv:1409.3465 [hep-th]}}.
\bibitem{TABruzzo:2014jza}
U.~Bruzzo, F.~Sala, and R.~J. Szabo, ``{${\mathcal{N} = 2}$ Quiver Gauge
  Theories on A-type ALE Spaces},''
  \href{http://dx.doi.org/10.1007/s11005-014-0734-x}{{\em Lett. Math. Phys.}
  {\bf 105} (2015) no.~3, 401--445},
\href{http://arxiv.org/abs/1410.2742}{{\tt arXiv:1410.2742 [hep-th]}}.
\bibitem{ContributionDI}
T.~Dimofte, ``Perturbative and nonperturbative aspects of complex Chern-Simons
  Theory,'' {\em Journal of Physics A} {\bf xx} (2016)  000,
  \href{http://arxiv.org/abs/1608.02961}{{\tt 1608.02961}}.

\end{thebibliography}\endgroup

\fi
}

\newcommand{\documentfinishBBL}{
\addcontentsline{toc}{section}{References}
\ifx\ifLONG\undefined
\input{\IDvalue.separate.bbl}
\end{document}
\else
\input{\DIRvalue/\IDvalue.volume.bbl}
\fi
}

\ifx\ifLONG\undefined
\def\volcite#1{Contribution \cite{Contribution#1}}
\else 
\def\volcite#1{Chapter \ref{Chapter#1}}
\fi

\usepackage{graphicx}
\usepackage{amsmath}
\usepackage{amssymb}
\usepackage{cite}
\usepackage{hyperref}
\usepackage{tikz}
\usepackage{geometry}
\usepackage{fullpage}

\numberwithin{equation}{section}

\newcommand{\BC}{\mathbb{C}}
\newcommand{\BR}{\mathbb{R}}
\newcommand{\BZ}{\mathbb{Z}}
\newcommand{\vev}[1]{\langle#1\rangle}
\newcommand{\tr}{\mathop{\mathrm{tr}}\nolimits}
\newcommand{\diag}{{\mathop{\mathrm{diag}}\nolimits}}

\newcommand{\CalH}{\mathcal{H}}
\newcommand{\CalS}{\mathcal{S}}
\newcommand{\CalM}{\mathcal{M}}
\newcommand{\CalL}{\mathcal{L}}

\begin{document}
\thispagestyle{empty}
\documentheader
\else \chapterheader \fi

\newcommand{\TAinc}[1]{\vcenter{\hbox{\includegraphics[scale=.25]{TA#1}}}}
\newcommand{\Nequals}[1]{$\mathcal{N}{=}#1$}

\newcommand{\TASL}{\mathrm{SL}}
\newcommand{\TASU}{\mathrm{SU}}
\newcommand{\TAU}{\mathrm{U}}
\newcommand{\TASO}{\mathrm{SO}}

\newcommand{\TAsg}{\mathsf{g}}

\numberwithin{equation}{section}

\section{Introduction}\label{sec:intro}
The aim of this review article is to give an elementary account of the 2d/4d correspondence, originally found in \cite{TAAlday:2009aq,TAGadde:2011ik}.  Let us begin by presenting  the essential idea, which is in fact quite simple.

We start from a certain six-dimensional quantum field theory $\CalS$, and consider its partition function on a product manifold $X_4\times C_2$, where $X_4$ is four-dimensional and $C_2$ is two-dimensional. 
Let us further suppose that thanks to the supersymmetric twists, the resulting partition function depends on the shapes but not on the sizes of $X_4$ and $C_2$. Then the six-dimensional partition function $\CalS(X_4\times C_2)$ can be evaluated in two ways. On the one hand, if we make $C_2$ very small, we first have a four-dimensional theory $\CalS(C_2)$, and then we can consider its partition function on $X_4$, namely $\CalS(C_2)(X_4)$.  On the other hand, if we make $X_4$ very small, we first have a two-dimensional theory $\CalS(X_4)$, and then we can consider its partition function on $C_2$, namely $\CalS(X_4)(C_2)$. 
We now have an equality \begin{equation}
\CalS(X_4)(C_2) = \CalS(C_2)(X_4).\label{fund-eq}
\end{equation}

So far the construction is extremely general. To get something concrete, we need to make a choice.  As an example, let us take $\CalS$ to be the 6d \Nequals{(2,0)} theory of type $A_1$, and $C_2$ to be an arbitrary two-dimensional surface. The 4d theory $\CalS(C_2)$ thus obtained is often called a class S theory of type $A_1$, and is an \Nequals{2} supersymmetric theory with a number of $\TASU(2)$ gauge group factors coupled to a number of trifundamental fields \cite{TAGaiotto:2009we}.   

If we choose $X_4=S^4$,  the partition function $\CalS(C_2)(X_4)$ can be computable by localization \cite{TANekrasov:2002qd,TAPestun:2007rz}, and from the results of the computation, one sees that $\CalS(S^4)$ is the Liouville theory, which is a non-compact 2d conformal field theory \cite{TAAlday:2009aq}.   If we choose $X_4=S^3\times S^1$, the partition function $\CalS(C_2)(X_4)$ is called the superconformal index \cite{TARomelsberger:2005eg,TAKinney:2005ej}, and from the results of the computation, one sees that $\CalS(S^3\times S^1)$ is the $q$-deformed $\TASU(2)$ Yang-Mills theory \cite{TAGadde:2011ik,TAGadde:2011uv}. 
These are the simplest cases of the correspondences, and various generalizations are possible and have been carried out. 

It is already five years since these correspondences were originally found\footnote{The bulk of this article was written in August 2014, with only a minor update on the references in the summer of 2016.}, and countless pages of original articles have been already wasted to describe the details and the generalizations.   The number of the review articles devoted to this topic is also already quite large. However, this huge amount of information can also be somewhat daunting, and the author therefore feels that it would be not completely useless to have another concise review, so that a newly interested reader  can quickly go through to have an idea of how this correspondence came to be known,  where the details can be learned,  and  what are still unsolved problems s/he might want to study.

This article is therefore intentionally meant to be a shallow overview.  Many of the facts will be stated as facts and will not be explained. Details and subtleties will be mentioned but will not be treated in full; references to review or original articles will be given instead.\footnote{The references are not at all exhaustive, and not even extensive either. The author will happily include more in the arXiv version, so please do not hesitate to email him. } 
The presentation will not be completely logical either.  It would be most systematic to start from six dimensions, to analyze the compactification very carefully, and to arrive at the  correspondence at the last step.  Thanks to the recent developments,  it would not be impossible to write a review in this order.  This will, however, be a hard read for people new to this field. 

Instead, this review will be organized to explain \emph{how} the correspondence works instead of \emph{why} there is the correspondence. In Sec.~\ref{sec:2}, we begin by learning two two-dimensional field theories that will be important for us: the two-dimensional $q$-deformed Yang-Mills theory and the Liouville theory. 
In Sec.~\ref{sec:4}, we introduce the class S theories of type $\TASU(2)$, directly as four-dimensional field theories defined by Lagrangian associated to Riemann surfaces with decompositions to three-punctured spheres. 
In Sec.~\ref{sec:part}, we quickly introduce the technique of  supersymmetric localization, and 
describe how the partition functions on $S^1\times S^3$ or on $S^4$ can be computed. 
In Sec.~\ref{sec:correspondences}, we apply the supersymmetric localization to the class S theories of type $\TASU(2)$. We will see that  the partition functions  on $S^3\times S^1$ and on $S^4$ are given by the $q$-deformed Yang-Mills and by the Liouville theory, respectively. 
We then explain how this correspondence can be understood in terms of  the 6d \Nequals{(2,0)} theory.
We will conclude  in Sec.~\ref{sec:conclusions} by going over possible future directions. 

\section{Two two-dimensional theories}\label{sec:2}
\subsection{Two-dimensional Yang-Mills theories}\label{sec:2d-gauge}
Let us first study two-dimensional Yang-Mills theories. We will first deal with the standard undeformed gauge theories, and will indicate how it can be $q$-deformed at the end. Every detail of the undeformed theory can be found in the great review \cite{TACordes:1994fc}. 

\subsubsection{Action}
 The 2d Yang-Mills theory with the gauge group $G$ has the Lagrangian  \begin{equation}
S\propto \frac{1}{e^2} \int d^2 x \sqrt{\det \TAsg} \tr F_{\mu\nu} F^{\mu\nu}.\label{2dgaugeaction}
\end{equation} Here we consider the theory on a curved manifold with the background metric $\TAsg$, in the Euclidean signature. The coupling constant $e$  can be removed  by rescaling $\TAsg$.  

Recall that in two dimensions, the only nonzero component of the field strength $F_{\mu\nu}$ is $F_{01}$.  
The kinetic term can then be written as \begin{equation}
\tr F_{\mu\nu}F^{\mu\nu}\propto (\TAsg^{00}\TAsg^{11}-\TAsg^{01}\TAsg^{10}) \tr (F_{01})^2 = (\det \TAsg)^{-1} \tr (F_{01})^2.
\end{equation}
This means that in the action \eqref{2dgaugeaction} we do not have individual components of the metric $\TAsg$: the only combination that appears is $\det \TAsg$. 
Put differently, the 2d Yang-Mills theory can be formulated on a 2d surface not quite equipped with the metric which allows us to measure the distance; all what we need is the volume form $dx^0 dx^1 \sqrt{\det \TAsg}$ which allows us to measure the area.  The only invariant of the 2d surface is then its genus and the total area of the surface, on which alone the partition function can depend. 

\subsubsection{On the cylinder}
\begin{figure}[h]
\[
\TAinc{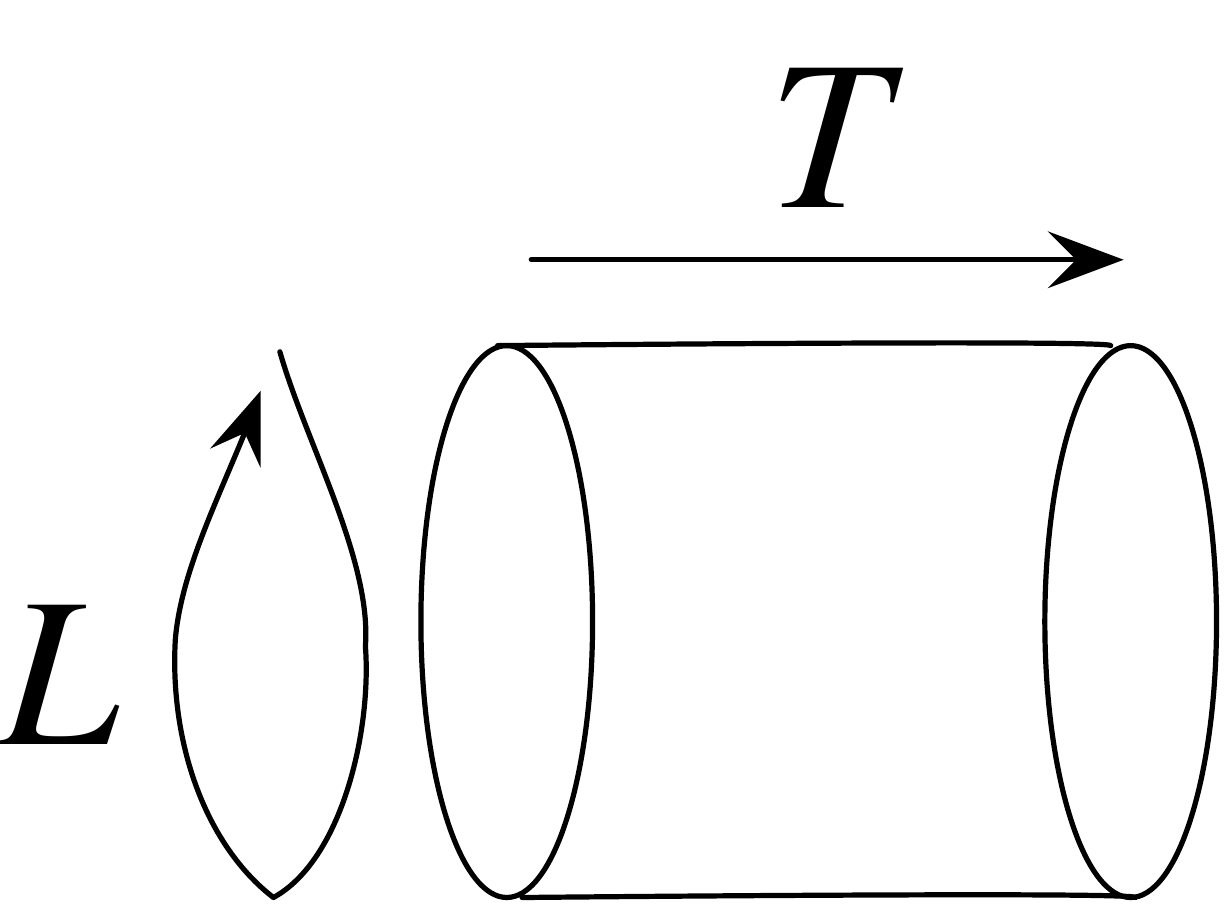}
\]
\caption{A cylinder with circumference $L$ \label{fig:cylinder}}
\end{figure}

Let us now analyze the theory on a cylinder $x^1\sim x^1+L$ with $x^0$ as the time direction, see Fig.~\ref{fig:cylinder}. We take the temporal gauge $A_0=0$.   At a constant time slice $x^0=0$, the gauge-invariant data is the holonomy of the gauge field around the circle: \begin{equation}
U:=P\exp\int_0^{L} A_1 dx^1\in G,
\end{equation} considered up to the adjoint $G$-action $U \mapsto g U g^{-1}$.

Then the wavefunction of the system is a function $\psi(U)$ defined on $G$, such that we have the invariance \begin{equation}
\psi(U)=\psi(gUg^{-1}).
\end{equation} It is a standard fact in group theory that such functions are given by a linear combination of traces \begin{equation}
\chi_R(U)=\tr_R U
\end{equation} in irreducible representations $R$.
Note that they are orthonormal under the natural measure on $G$: \begin{equation}
\int_G \chi_R (U) \chi_{R'}( U)^* dU = \delta_{RR'}.\label{grouportho}
\end{equation}

Decomposing $A_1=A_1^a T_a$, where $a=1,\ldots,\dim G$, the Hamiltonian obtained from \eqref{2dgaugeaction} is \begin{equation}
H\propto \int_0^L \frac{\delta}{\delta A_1^a(x)}\frac{\delta }{\delta A_1^a(x)} dx^1.
\end{equation} Acting on $\chi_R(U)=\tr_R P\exp\int_0^L A_1 dx^1 $, we find \begin{equation}
H\chi_R(U)\propto \tr_R \int_0^L T^a T^a dx^1 P\exp(\int_0^L A_1dx^1) \propto L c_2(R) \chi_R(U)
\end{equation} where $c_2(R)$ is the value of the quadratic Casimir in the irreducible representation $R$.
We fix the proportionality constants by demanding that $H\chi_R=Lc_2(R)\chi_R$. 

Now we can evaluate the partition function $Z$ on a torus $x^1\sim x^1+L$, $x^0\sim x^1+T$: \begin{equation}
Z=\tr e^{-TH} = \sum_R e^{-TL c_2(R)}.
\end{equation} Note that the final result only depends on the total area $TL$ of the torus, as it should be.

\subsubsection{On a general surface}

\begin{figure}[h]
\[
\TAinc{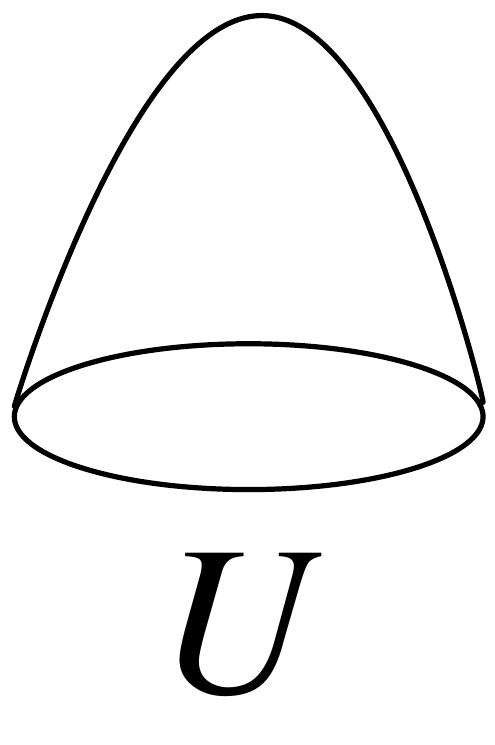}\qquad\TAinc{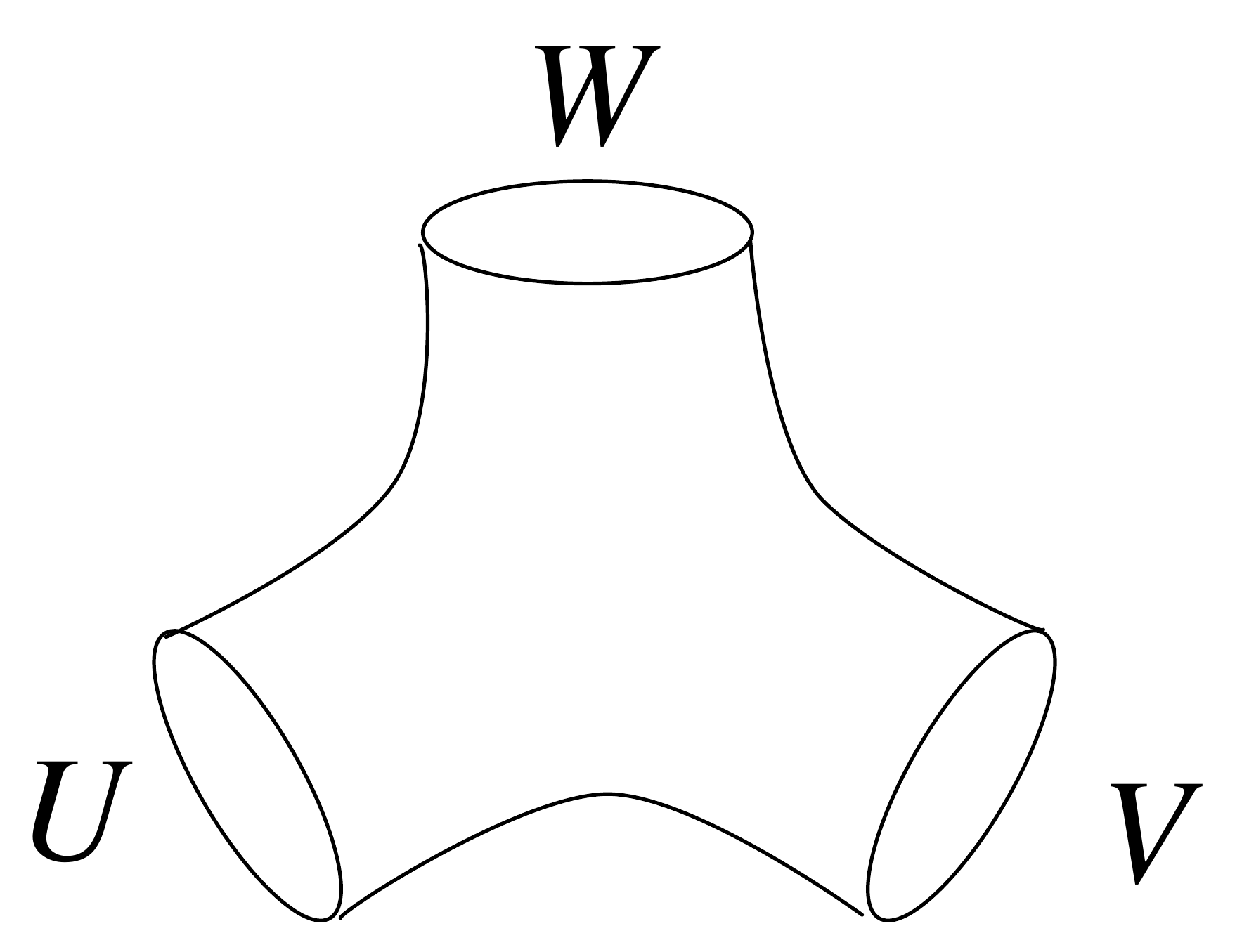}
\]
\caption{A disk and a three-holed sphere, with holonomies around the boundaries specified \label{fig:disk}}
\end{figure}

Next, let us study the theory on a general 2d surface. First, consider the case of a disk with area $A$, see Fig.~\ref{fig:disk}. 
We specify the holonomy $U$ around the boundary circle; then we can perform the partition function under this condition. We can denote it as $Z_A(U)$.  This can also be thought of as defining a wavefunction on the boundary $S^1$, and can be denoted as $\psi_A(U)=Z_A(U)$.  Therefore,  in the following, we use the terminology the partition function $Z$ and the wavefunction $\psi$ interchangeably.  This method of defining a wavefunction via a path-integral over a disk (or more generally a ball in higher dimensions) was pionneered by Hartle and Hawking, and therefore this is often called the Hartle-Hawking wavefunction.    

\begin{figure}[h]
\[
\TAinc{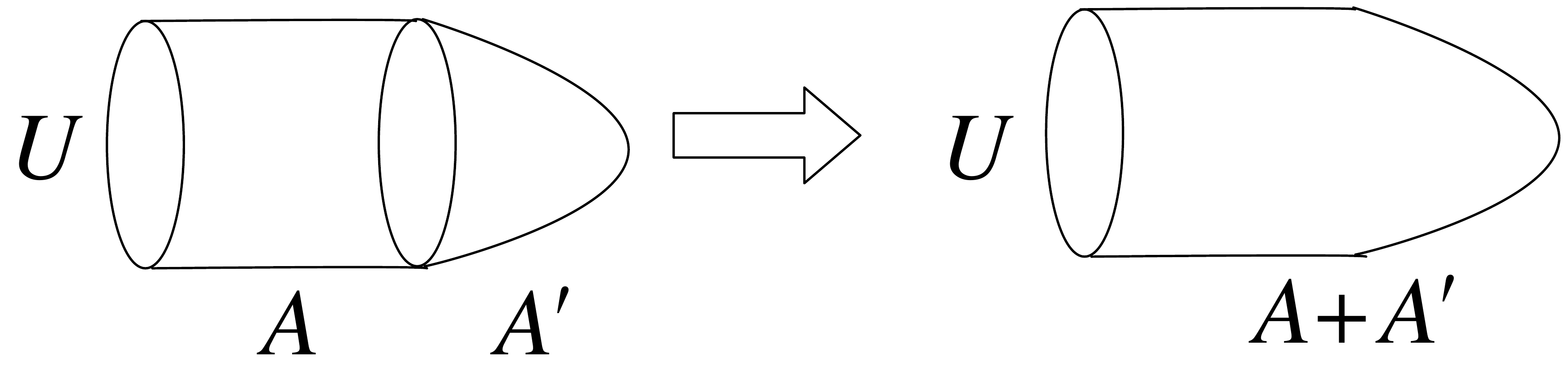}
\]
\caption{The area of a disk can be changed by gluing a cylinder\label{fig:add}}
\end{figure}

So, what is this wavefunction $\psi_A(U)$ associated to the disk?
First, note that we can glue a cylinder of area $A$ to a disk of area $A'$ to have a disk of area $A+A'$, see Fig.~\ref{fig:add}.  This means \begin{equation}
\psi_{A+A'}(U)= e^{-Ac_2}\psi_{A'}(U)
\end{equation} where $c_2$ is the operator acting by $c_2(R)$ on $\chi_R(U)$.
Therefore it suffices to determine $\psi_{A=0}(U)$. 
When the area is zero, $U$ is forced to be an identity element, and therefore $\psi_{A=0}(U)=\alpha\delta(U)$
where $\delta(U)$ is the delta function at the identity on the group manifold of the group $G$ and $\alpha$ is a proportionality constant.
We can write \begin{equation}
\delta(U)=\sum_R d_R \chi_R(U)
\end{equation} where $d_R$ can be found from the orthonormality property \eqref{grouportho}: \begin{equation}
d_R=\int_G \delta(U) \chi_R(U) dU = \tr_R 1 = \dim R.
\end{equation} This way we find \begin{equation}
\psi_A(U)= \alpha\sum_R e^{-A c_2(R)} (\dim R) \chi_R(U).\label{disk}
\end{equation}

Second, it is useful at this point to rewrite the Hamiltonian on the cylinder we found above as the amplitude on the cylinder whose boundary holonomies are $U$, $V$: \begin{equation}
\psi_A(U,V)=\sum_R e^{-A c_2(R)} \chi_R(U) \chi_R(V^{-1}).\label{cylinder}
\end{equation}

\begin{figure}[h]
\[
\TAinc{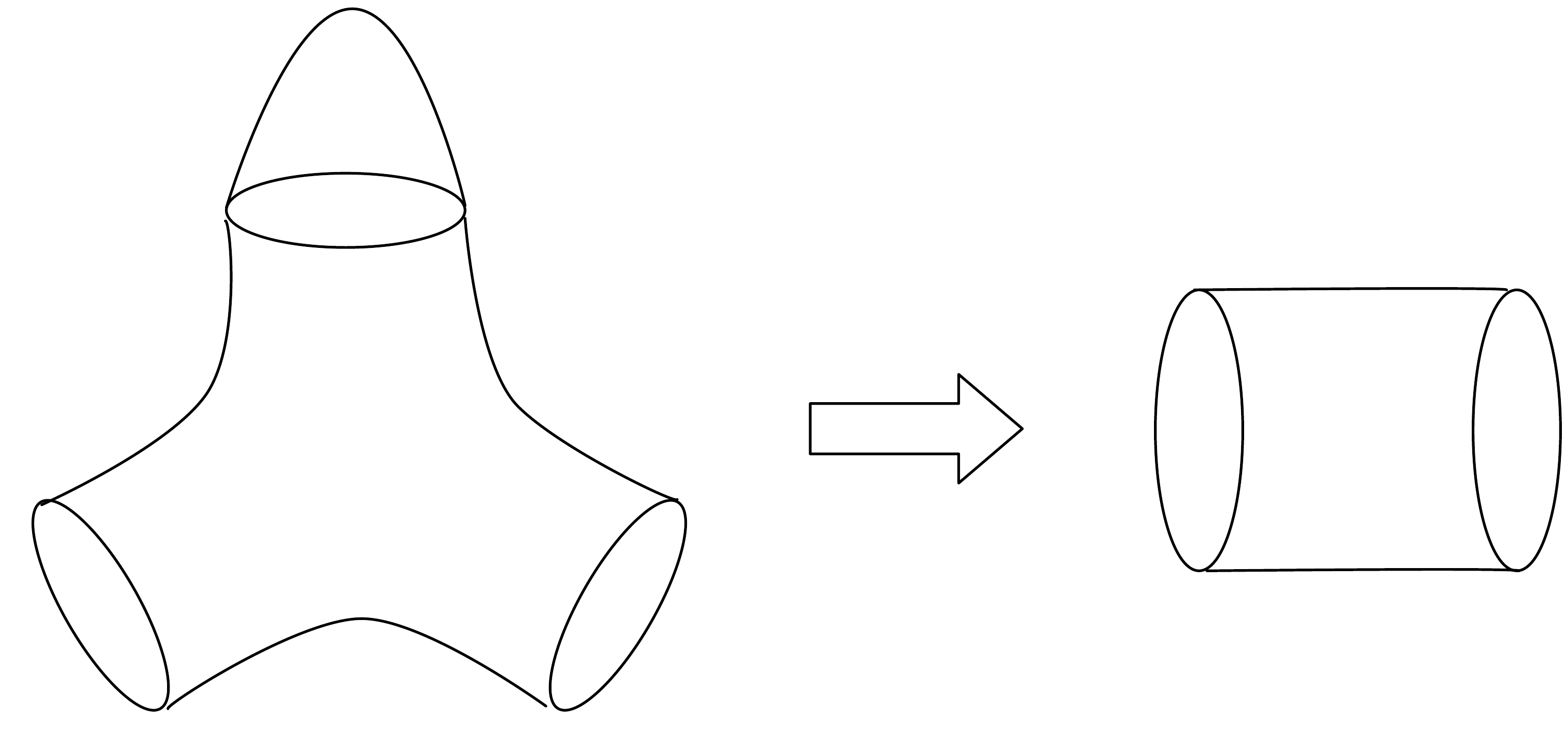}
\]
\caption{Sewing a disk to a hole makes a three-holed sphere into a cylinder \label{fig:attach}}
\end{figure}

Third, note that any 2d surface can be cut into pieces, such that each piece is a sphere with three holes, see Fig.~\ref{fig:disk}.
Let us say the area is $A$ and the holonomies around the three holes are $U$, $V$ and $W$. What is the wavefunction $\psi_A(U,V,W)$?  The crucial property is that 
when a disk is sewed to a hole, it becomes a cylinder, see Fig.~\ref{fig:attach}. In terms of an equation, this becomes 
\begin{equation}
\int_G \psi_A(U,V,W) \psi_{A'}(W^{-1}) dW = \psi_A(U,V).
\end{equation} 

 Using \eqref{disk} and \eqref{cylinder}, one finds that the unique solution is \begin{equation}
\psi_A(U,V,W)=\frac{1}{\alpha}\sum_R e^{-Ac_2(R)} (\dim R)^{-1} \chi_R(U) \chi_R(V) \chi_R(W).
\end{equation}

\begin{figure}[h]
\[
\TAinc{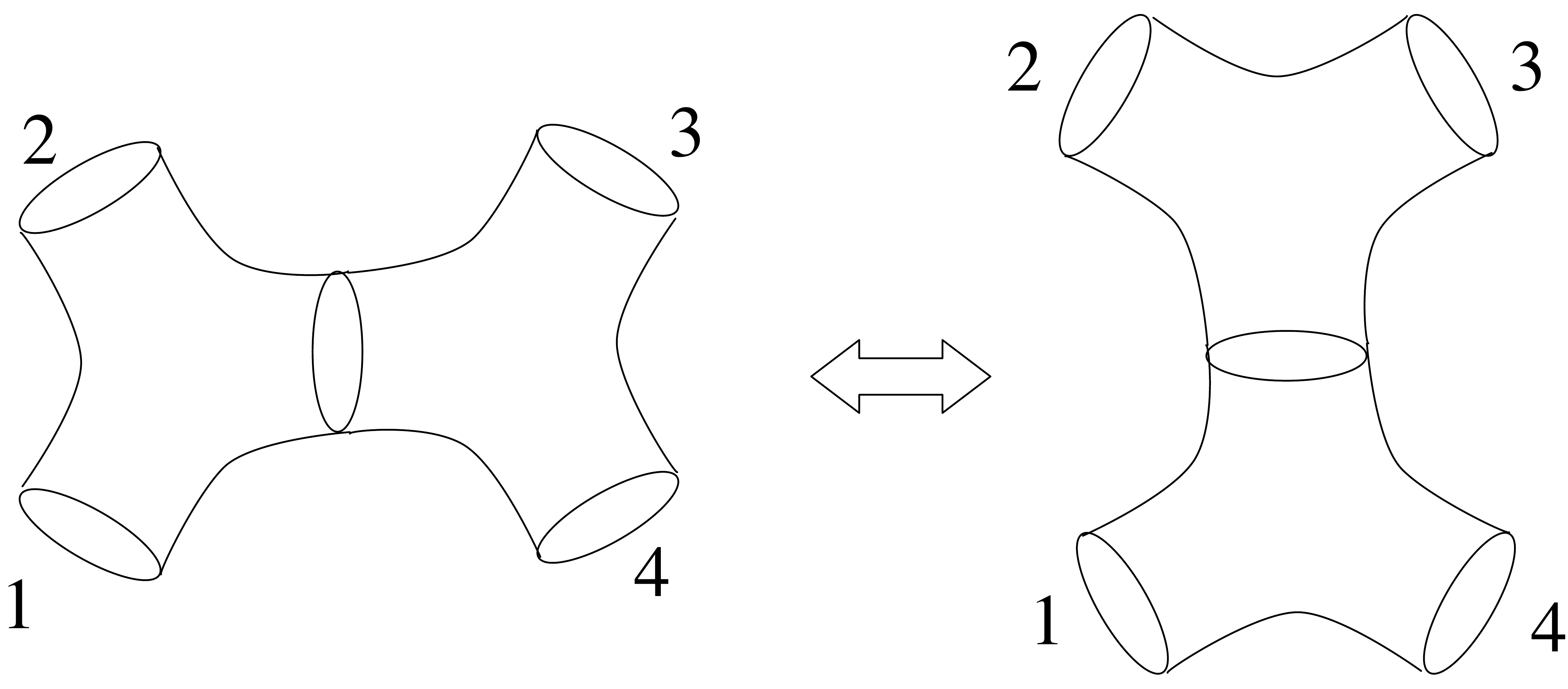}
\]
\caption{The amplitude does not depend on how one cuts a four-holed sphere into two three-holed spheres \label{fig:assoc}}
\end{figure}

The result can be easily generalized to arbitrary surface of area $A$, genus $g$ and $n$ holes with holonomies $U_{i=1,\ldots,n}$, by gluing $\psi_A(U,V,W)$. The answer is \begin{equation}
\psi_{A,g}(U_i) = \alpha^{2-2g-n}\sum_R e^{-A c_2(R)} \frac{\prod_i \chi_R (U_i)}{(\dim R)^{2g-2+n}}.\label{2dgaugefinal}
\end{equation}
Note that this final answer automatically satisfies the associativity of the sewing of two three-punctured spheres as shown in Fig.~\ref{fig:assoc}. 

Finally let us briefly discuss the dependence on $\alpha$ of various quantities found above. 
In the action \eqref{2dgaugeaction} we can include a local term \begin{equation}
\delta S= \beta\int d^2 x \sqrt{\TAsg} R \label{chiR}
\end{equation} where $R$ is the Riemann curvature of the metric $\TAsg$. On a surface without punctures, this integrates to $\beta(2-2g)$, and is therefore topological.  When the surface has punctures, there is a natural boundary term that makes the integral $\beta(2-2g-n)$.   
By including this local but topological term \eqref{chiR}, the factors $\alpha^{2-2g-n}$ we saw above shifts to $(e^\beta \alpha)^{2-2g-n}$.  

A term of the form \eqref{chiR} can be easily generated by changing the regularization scheme and/or the renormalization scheme of a quantum field theory. Therefore there is not much sense in asking what value of $\alpha$ we get when we start from $\beta=0$  in the original Lagrangian \eqref{2dgaugeaction}.

\subsubsection{$q$-deformation}
So far, we solved the 2d Yang-Mills theory starting from the action \eqref{2dgaugeaction}.
We can instead start from the lattice formulation. Namely, we draw a sufficiently fine mesh on the 2d surface.
At each edge $e$, we assign a dynamical variable $U_e$ taking values in the group $G$, and at each face $f$, we assign a Boltzmann  weight \begin{equation}
\psi_{f}=\sum_R e^{-A_f c_2(R)}\tr_R \prod_{e}  U_e
\end{equation} where the product is taken around the edges  $e$ around the face $f$.  Then the path integral defined as \begin{equation}
Z=\int  \prod_e dU_e  \prod_f \psi_{f}  \label{latticeintegral}
\end{equation} gives the partition function \eqref{2dgaugefinal}.

An interesting deformation of this theory is obtained by declaring that edge variables $U_e$ take values in the quantum group $G_q$, instead of in the ordinary group $G$.  The quantum groups are obtained by making non-commutative the matrix entries of the group. For example, the quantum group $\TASU(2)_q$ is given by considering $2\times 2$ matrices $U_i{}^j=\begin{pmatrix}
\alpha & \beta\\
\gamma & \delta
\end{pmatrix}$ with the relations
\begin{align}
\alpha\beta&=q^{1/2}\beta\alpha, & \alpha\gamma&=q^{1/2}\gamma\alpha, &
\beta\delta &=q^{1/2}\delta\beta, & \gamma \delta &=q^{1/2}\delta\gamma, &
\beta\gamma&=\gamma\beta
\end{align}
and \begin{equation}
\alpha \delta - q^{1/2}\gamma \beta= \delta \alpha-q^{-1/2}\gamma \beta=1,
\end{equation}
with their complex conjugates given by 
\begin{equation}
\bar U_{\bar\imath }^{\bar\jmath }=\begin{pmatrix}
\alpha^* & \beta^*\\
\gamma^* & \delta^*
\end{pmatrix}=
\begin{pmatrix}
\delta & -q^{1/2}\gamma \\
-q^{-1/2}\beta & \alpha
\end{pmatrix}.
\end{equation}
Since the matrix entries themselves are non-commutative, it is slightly tricky to come up with a correct ordering of variables in the lattice path integral \eqref{latticeintegral} but this can be done \cite{TABuffenoir:1994fh,TAAlekseev:1994au}. 

Let us see how a complication would arise, in a simple example. 
From the explicit commutation relations of the matrix entries of $\TASU_q(2)$ given above, it is easy to check that we have \begin{equation}
\delta_{j\bar\jmath }U_i{}^j \bar U_{\bar\imath }{}^{\bar\jmath }=\delta_{i\bar\imath }.
\end{equation} 
However, due to the non-commutativity of the entries, we have \begin{equation}
\delta^{i\bar\imath }U_i{}^j \bar U_{\bar\imath }{}^{\bar\jmath } \neq \delta^{j\bar\jmath }.
\end{equation} 
Instead, we have \begin{equation}
D^{i\bar\imath }U_i{}^j \bar U_{\bar\imath }{}^{\bar\jmath } = D^{j\bar\jmath } 
\end{equation} where \begin{equation}
D^{i\bar\imath }=\begin{pmatrix}
q^{-1/2} & 0 \\
0 & q^{1/2}
\end{pmatrix}.
\end{equation} 
Therefore, the natural combination $\delta^{i\bar\imath}\delta_{i\bar\imath}$ of the undeformed $\TASU(2)$ is modified to 
\begin{equation}
\delta^{i\bar\imath}\delta_{i\bar\imath}=2 \quad\rightsquigarrow\quad D^{i\bar\imath}\delta_{i\bar\imath}=q^{1/2}+q^{-1/2}
\end{equation}
in the representation theory of $\TASU_q(2)$.
The right hand side is called the quantum dimension of the two-dimensional representation of $\TASU_q(2)$.

At the end of the day, the only change in the final expression \eqref{2dgaugefinal} of the partition function, due to the fact that the gauge group is now the quantum group, is that the dimension $\dim R$ is replaced by the quantum dimension $\dim_q R$.
For the general $\TASU(N)$ case, the quantum dimension is given by \begin{equation}
\dim_q R= \tr_R \diag(q^{(N-1)/2},q^{(N-3)/2},\ldots, q^{(1-N)/2}),
\end{equation}  and therefore  the partition function of the $q$-deformed theory is \begin{equation}
\psi_{A,g}(U_i) = \alpha^{2-2g-n}\sum_R e^{-A c_2(R)} \frac{\prod_i \chi_R (U_i)}{(\dim_q R)^{2g-2+n}}.\label{2dqdeformedfinal}
\end{equation} 
It is known that the same deformation arises also string theoretically \cite{TAAganagic:2004js}, although the underlying quantum group is not directly visible there.

\subsection{The Liouville theory}\label{sec:2d-liou}
We will now study the second two-dimensional field theory, known as the Liouville theory. It is the prime example of so-called irrational conformal field theory.  
Here we cover only extremely shallow aspects of this beautiful and rich theory. An interested reader is referred to the classic reviews such as \cite{TASeiberg:1990eb,TANakayama:2004vk}. We will mainly use a more axiomatic approach, pioneered and reviewed in \cite{TATeschner:2001rv}.
Before getting there, let us quickly recall the free boson theory. 

\subsubsection{Free boson theory}
A massless boson $\phi$ in two dimensions in the Euclidean signature satisfies the equation of motion \begin{equation}
\triangle \phi=\bar\partial \partial \phi =0.
\end{equation} A general solution is then \begin{equation}
\phi(z,\bar z)= f(z)+\bar f(\bar z)
\end{equation} where $f(z)$ is a holomorphic function. 

From this we see that the classical theory has a symmetry under arbitrary holomorphic changes of the coordinate $z\mapsto g(z)$. 
Denote the generator of the infinitesimal transformation $z\to z(1+\epsilon z^n)$ by $L_n$. Classically they satisfy the commutation relation $[L_m,L_n]=(m-n)L_{m+n}$. 

In the quantum theory, the basic operator product expansion of the free boson theory is \begin{equation}
\partial\phi(z) \partial \phi(w)\sim -\frac{1}{2(z-w)^2}
\end{equation} and \begin{equation}
e^{2ia\phi(z,\bar z)}e^{-2ia\phi(w,\bar w)} \sim \frac{1}{|z-w|^{2a^2}}.\label{exp}
\end{equation}
The generator $L_0$ rescales the coordinate as $z \mapsto e^{-\epsilon L_0} z$. Correspondingly, when the two-point function behaves as $(z-w)^{-2\Delta}(\bar z-\bar w)^{-2\bar\Delta}$, we say that the operator has the holomorphic dimension $L_0=\Delta$ and the anti-holomorphic dimension $\bar L_0=\bar\Delta$. We find $\partial\phi$ has $(L_0,\bar L_0)=(1,0)$ and $e^{2ia\phi}$ has $(L_0,\bar L_0)=(a^2,a^2)$.

Quantum mechanically, the algebra generated by $L_n$ is modified to \begin{equation}
[L_m,L_n]=(m-n) L_{m+n} + c\frac{m^3-m}{12}\delta_{m,-n}
\end{equation} where $c$ is a number called the central charge. This is the celebrated Virasoro algebra.  We package them into a field $T(z)=\sum L_n z^{n-2}$. The commutation relation above is equivalent to the operator product expansion \begin{equation}
T(z)T(w) = \frac{c}{2}\frac{1}{(z-w)^4}+\frac{2}{(z-w)^2} T(w)+\frac{1}{z-w} T'(w) + \cdots.\label{tt}
\end{equation}

In a free-boson theory, the energy-momentum tensor $T(z)$ is given by $T(z)=(1/2)\partial\phi \partial\phi$. A short computation reveals that it satisfies the relation \eqref{tt} with $c=1$.  We now consider a slightly modified free-boson theory where the energy momentum tensor is given by \begin{equation}
T(z)=\partial\phi \partial\phi + i Q \partial^2\phi.
\end{equation} We find that this satisfies the relation \eqref{tt} with \begin{equation}
c=1+ 6 Q^2.
\end{equation} This modification corresponds to having the Lagrangian density \begin{equation}
\CalL= g^{ij}\partial_i \phi \partial_j \phi + Q\phi R\label{lineardilaton}
\end{equation}
 where $R$ is the curvature of the 2d surface.  When computing the correlators on the sphere, we can map it to a flat infinite plane, with the caveat that there is still a concentration of the curvature at $z=\infty$. This effectively place an operator $e^{-2Q\phi}$ at $z=\infty$. 
From this reason the parameter $Q$ is often called the background charge. 
This modifies the basic correlator of the exponential fields \eqref{exp} to \begin{equation}
e^{2(Q/2+ia)\phi(z,\bar z)}e^{2(Q/2-ia)\phi(w,\bar w)} \sim  \frac{1}{|z-w|^{2a^2+Q^2/2}}.
\end{equation} In particular, the exponential operator  $e^{2(Q/2+ia)\phi}$ has $L_0=a^2+Q^2/4$.
More generally, we say that the operator $e^{2a\phi}$ has \begin{equation}
L_0=a(Q-a).\label{lineardilatondim}
\end{equation}

\subsubsection{Interacting theory}
Suppose now that we want to change this free boson theory with background charge $Q$ into a full-fledged interacting theory without destroying the conformal invariance. At leading order, a new term in the Lagrangian should have the dimension $(L_0,\bar L_0)=(1,1)$ to preserve the conformal invariance. 
From \eqref{lineardilatondim}, we find that the operator $e^{2b\phi}$ does the job, when \begin{equation}
Q=b+\frac1b.
\end{equation} Now the Lagrangian density is \begin{equation}
\CalL= g^{ij}\partial_i \phi \partial_j \phi + Q\phi R+ 4\pi \mu e^{2b\phi }\label{liouvillelag}
\end{equation} where the parameter $\mu$ is often called the cosmological constant.

This parameter $\mu$  can be set to any value one wants, by shifting the origin of $\phi$, so it is not easy to do a perturbation theory in terms of this interaction term.
Put differently, the potential term $e^{2b\phi}$ is exponential, and cannot be considered as a small deformation from the free theory with the Lagrangian \eqref{lineardilaton}. But after a series of impressive works, we now know that the Lagrangian density \eqref{liouvillelag} determines an interacting conformal field theory with the central charge \begin{equation}
c=1+6Q^2=1+6(b+\frac1b)^2.\label{liouvillecentral}
\end{equation}

One crucial difference from the free theory is as follows.  In the free theory, the operator $e^{2ip\phi}$ with $L_0=p^2$ gives rise to a state with momentum $p$ moving the direction parameterized by $\phi$, under the state-operator correspondence. In particular, two operators $e^{2ip\phi}$ and $e^{-2ip\phi}$ are two distinct states.

Similarly, in the Liouville theory, the operator $e^{(Q+2ip)\phi}$ with $L_0=p^2+(Q/2)^2$ corresponds to a state  with momentum $p$ in the $\phi$-space. Here the shift of the exponent by $Q$ is necessary to keep $L_0$ real and positive.   Due to the exponential interaction $e^{2b\phi}$, the wave coming from the negative $\phi$ region cannot penetrate to the positive $\phi$ region. Instead, it gets reflected by the exponential potential wall.  This means that two operators $e^{(Q+2ip)\phi}$ and $e^{(Q-2ip)\phi}$ give one and the same state: \begin{equation}
e^{(Q+2ip)\phi}=R(p)e^{(Q-2ip)\phi}
\end{equation} where $R(p)$ is a phase called the reflection coefficient. The asymptotic behavior of $R(p)$ can be computed from the quantum mechanical scattering problem by the exponential potential, and the constraints on further corrections to $R(p)$ from the conformal invariance was one of the starting points of the full solution of the Liouville theory. 

We now know that a unitary conformal theory can be uniquely specified by the condition that its spectrum is given by a family of primary operators $V_p(z)$ for  a real number $p\ge 0$ with  $L_0=\bar L_0=p^2+(Q/2)^2$. The central charge is \eqref{liouvillecentral}. 
The three-point function  on the sphere was originally found independently by  \cite{TADorn:1994xn} and \cite{TAZamolodchikov:1995aa}. 
Using conformal invariance, we can put one operator at $z=\infty$, another at $z=1$, and the third at $z=0$.
Then it is given by
\begin{multline}
C(\alpha_1,\alpha_2,\alpha_3):=\vev{e^{2(Q-\alpha_1)\phi(\infty)}
e^{2\alpha_2\phi(1)}
e^{2\alpha_3\phi(0)}}
= 
\left[\pi \mu \gamma(b^2) b^{2-2b^2}\right]^{(Q-\alpha_1-\alpha_2-\alpha_3)/b} \\
\times \frac{\Upsilon'(0)\Upsilon(2\alpha_1)\Upsilon(2\alpha_2)\Upsilon(2\alpha_3)}
{\Upsilon(\alpha_1+\alpha_2+\alpha_3-Q)
\Upsilon(\alpha_1+\alpha_2-\alpha_3)
\Upsilon(\alpha_1-\alpha_2+\alpha_3)
\Upsilon(-\alpha_1+\alpha_2+\alpha_3)}. \label{DOZZ}
\end{multline} Here, 
 \begin{equation}
\gamma(x)= \Gamma(x)/\Gamma(1-x)
\end{equation}
and \begin{equation}
\Upsilon(x)=\frac{1}{\Gamma_2(x|b,b^{-1})\Gamma_2(Q-x|b,b^{-1})}\label{ups}
\end{equation} where $\Gamma_2(x|\epsilon_1,\epsilon_2)$ is Barnes' double Gamma function 
obtained by regularizing the infinite product 
\begin{equation}
\Gamma_2(x|\epsilon_1,\epsilon_2) \propto \prod_{m,n \ge 0}
\left(x+m\epsilon_1+n \epsilon_2\right)^{-1}.\label{infiniteproduct}
\end{equation}

We will need the following properties of the double Gamma function later in this review. 
First, $\Gamma_2(x|\epsilon_1,\epsilon_2)$  is real when $x$ is real, assuming $\epsilon_{1,2}$ are real. 
Then, from analytic continuation, we have \begin{equation}
\Gamma_2(x^*|\epsilon_1,\epsilon_2)=
\Gamma_2(x|\epsilon_1,\epsilon_2)^*.
\end{equation}
Another relation we need is \begin{equation}
\Gamma_2(x+\epsilon_1|\epsilon_1,\epsilon_2)\Gamma_2(x+\epsilon_2|\epsilon_1,\epsilon_2)
=x \Gamma_2(x|\epsilon_1,\epsilon_2)\Gamma_2(x+\epsilon_1+\epsilon_2|\epsilon_1,\epsilon_2).\label{gamma-shift}
\end{equation}

Before proceeding further, note that the two- and three-point functions are essentially invariant under the exchange $b\to b^{-1}$, as the first line in \eqref{DOZZ} can be absorbed into the definition of the primary operators.  This invariance under the inversion of $b$ is rather surprising from the point of view of the Lagrangian description using \eqref{liouvillelag}, as it cannot be seen classically at all.  The invariance played a crucial role when people first solved  the Liouville theory. 

\subsubsection{Four-point function: general structure}
In the 2d gauge theory case, the knowledge of the cylinder partition function and the partition function for the sphere with three holes were enough to compute the partition function on arbitrary surface with any number of holes. 
Similarly, also for the Liouville theory, the knowledge of the two-point function and the three-point function is sufficient to obtain the correlation function on arbitrary surface with any number of insertions of operators. 

\begin{figure}[h]
\[
\TAinc{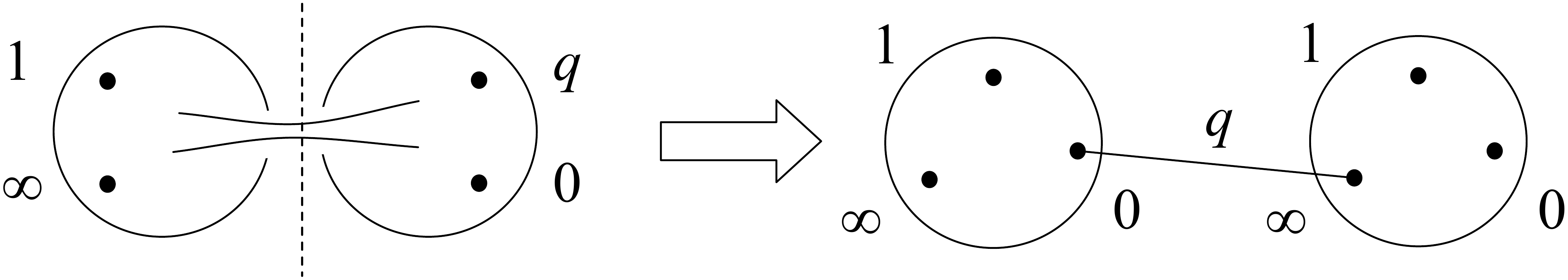}
\]
\caption{A four-punctured sphere is composed from two three-punctured sphere by gluing\label{fig:cut}}
\end{figure}

Let us illustrate the method by computing the four point function.
Using the conformal invariance, we can put three operators at $z=\infty$, $z=1$ and $z=0$.
Let $q$ be the position of the fourth operator, and we would like to obtain
 \begin{equation}
\vev{e^{2(Q-\alpha_1)\phi(\infty)}
e^{2\alpha_2\phi(1)}
e^{2\alpha_3\phi(q,\bar q)}
e^{2\alpha_4\phi(0)}}.
\end{equation} 
This is done by inserting the complete set of states between $e^{2\alpha_2\phi(1)}$
and $e^{2\alpha_3\phi(q,\bar q)}$, see Fig.~\ref{fig:cut}.

At the level of the geometry, we have a three-punctured sphere with  a local coordinate $z$, and another with a local coordinate $z'$. Both have three punctures at $z,z'=\infty,1,0$. 
We connect $z=0$ and $z'=\infty$; this is done as follows. The local coordinate at $z'=\infty$ is better thought of as $w=0$, where $wz'=1$.  Now, the gluing of two punctures with parameter $q$, one at $z=0$ and another at $w=0$ is done by performing the identification $zw=q$.  We end up having a four-punctured sphere with coordinate $z$, with punctures at $z=\infty,1,q$ and $0$.

  The complete set of states are given by the operators \begin{equation}
O_{p,\{n\},\{\tilde n\}}:= L_{-n_1}L_{-n_2} \cdots L_{-n_k} \bar L_{-\tilde n_1}\bar L_{-\tilde n_2} \cdots \bar  L_{-\tilde n_{\tilde k}} e^{2(Q/2+ip)\phi}
\end{equation}  where $p\ge 0$ is real and positive, and the positive integers \begin{equation}
n_1 \ge n_2 \ge \cdots \ge n_k,\quad
\tilde n_1 \ge \tilde n_2 \ge \cdots \ge \tilde n_{\tilde k}
\end{equation}  specify the descendants of the Virasoro algebra.  Note that $k$ or $\tilde k$ can be zero. 
The four-point function is then decomposed as \begin{multline}
\vev{e^{2(Q-\alpha_1)\phi(\infty)}
e^{2\alpha_2\phi(1)}
e^{2\alpha_3\phi(z,\bar z)}
e^{2\alpha_4\phi(0)}}=
\int_{0}^\infty dp \sum_{\{n\},\{\tilde n\},\{n'\},\{\tilde n'\}}
\vev{e^{2(Q-\alpha_1)\phi(\infty)}
e^{2\alpha_2\phi(1)}
O_{p,\{n\},\{\tilde n\}}}  \\
z^{-\sum n} \bar z^{-\sum \tilde n} |z|^{-2\Delta_{p}}
G^{\{n\},\{n'\}}
G^{\{\tilde n\},\{\tilde n'\}}
\vev{
O_{p,\{n'\},\{\tilde n'\}}{}^\dagger
e^{2\alpha_3\phi(1)}
e^{2\alpha_4\phi(0)}} \label{complete}
\end{multline} where 
$\Delta_p=p^2+Q^2/4$ and
$G_p^{\{n\},\{n'\}}G_p^{\{\tilde n\},\{\tilde n'\}}$ is the inverse matrix of \begin{multline}
G_{p,\{n\},\{n'\}}G_{p,\{\tilde n\},\{\tilde n'\}}
= \vev{O_{p,\{n'\},\{\tilde n'\}}{}^\dagger O_{p,\{n\},\{\tilde n\}}}\\
= \vev{e^{(Q-2ip)\phi(\infty)}
L_{n'_1} \cdots L_{n'_{k'}} \bar L_{\tilde n'_1} \cdots \bar  L_{\tilde n'_{\tilde k'}}
L_{-n_1} \cdots L_{-n_k} \bar L_{-\tilde n_1} \cdots \bar  L_{-\tilde n_{\tilde k}}
e^{(Q+2ip)\phi(0)}}.
\end{multline}  We will soon see that $G_{p,\{n\},\{n'\}}$ can be computed using only the Virasoro algebra.

We can also write  \begin{align}
\vev{e^{2(Q-\alpha_1)\phi(\infty)}
e^{2\alpha_2\phi(1)}
O_{p,\{n\},\{\tilde n\}}} 
&=
l_{\alpha_1,\alpha_2,\alpha,\{n\}}
l_{\alpha_1,\alpha_2,\alpha,\{\tilde n\}}
 C(\alpha_1,\alpha_2,\alpha) \\
\vev{
O_{p,\{n'\},\{\tilde n'\}}{}^\dagger
e^{2\alpha_3\phi(1)}
e^{2\alpha_4\phi(0)}
} 
&=
r_{\{ n'\},Q-\alpha,\alpha_3,\alpha_4}
r_{\{\tilde n'\},Q-\alpha,\alpha_3,\alpha_4}
 C(Q-\alpha,\alpha_3,\alpha_4)
\end{align}
where $\alpha=Q/2+ip$, and the functions $l$ and $r$ can again be computed using only the Virasoro algebra.

 Plugging these relations back in \eqref{complete}, we have \begin{multline}
\vev{e^{2(Q-\alpha_1)\phi(\infty)}
e^{2\alpha_2\phi(1)}
e^{2\alpha_3\phi(q,\bar q)}
e^{2\alpha_4\phi(0)}}=
\int_{0}^\infty dp  |q|^{2\Delta_p}\\
C(\alpha_1,\alpha_2,\frac Q2+ip)
C(\frac Q2-ip,\alpha_3,\alpha_4)
F_{\alpha_1,\alpha_2,Q/2+ip,\alpha_3,\alpha_4}(q)
\overline{F_{\alpha_1,\alpha_2,Q/2+ip,\alpha_3,\alpha_4}(q)}\label{liouville-four}
\end{multline}  where \begin{equation}
F_{\alpha_1,\alpha_2,Q/2+ip,\alpha_3,\alpha_4}(q) 
=
\sum_{\{n\},\{n'\}}q^{ \sum n}  l_{\alpha_1,\alpha_2,Q/2+ip,\{n\}}
G_p^{\{n\},\{n'\}} 
r_{\{\tilde n'\},Q/2-ip,\alpha_3,\alpha_4}
\end{equation} is known as the four-point conformal block.

Note that the expression \eqref{liouville-four} was obtained by inserting a complete set of states 
between pairs $\alpha_1$, $\alpha_2$ at $z=\infty,1$ and  $\alpha_3,\alpha_4$ at $z=q,0$.
The same correlator can also be obtained by inserting a complete set of states between 
pairs $\alpha_1$, $\alpha_3$ at $z=\infty,q$ and  $\alpha_2,\alpha_4$ at $z=1,0$ or between
pairs $\alpha_1$, $\alpha_4$ at $z=\infty,0$ and  $\alpha_2,\alpha_3$ at $z=1,q$.
The equality of the resulting expressions is not at all trivial, but has been proved in \cite{TAPonsot:1999uf,TATeschner:2001rv}.

\subsubsection{Four-point function: explicit expressions}
Let us determine the conformal block explicitly to the first few orders.
For the zeroth order term, we just have $G_{p,0,0}=l_{\alpha_1,\alpha_2,Q/2+ip,0}=r_{Q/2-ip,\alpha_3,\alpha_4,0}=1$, and so $F_{\alpha_1,\alpha_2,Q/2+ip,\alpha_3,\alpha_4}(q) = 1 +O(q)$. 

In the next order, we first compute \begin{equation}
\vev{e^{(Q-2ip)\phi(\infty)}L_1L_{-1} e^{(Q+2ip)\phi(0)}}
\end{equation} using the commutation relation \begin{equation}
L_1L_{-1}=L_{-1}L_1+ 2L_0,
\end{equation} and the fact $L_1$ annihilates the primary $e^{(Q+2ip)\phi}$.
As $L_0$ is $\Delta_p=p^2+(Q/2)^2$, we find \begin{equation}
G_{p,\{1\},\{1\}}=2\Delta_p.
\end{equation} At the next order, we find \begin{equation}
\begin{pmatrix}
G_{p,\{2\},\{2\}} & G_{p,\{2\},\{1,1\} } \\
G_{p,\{1,1\},\{2\}} & G_{p,\{1,1\},\{1,1\}}
\end{pmatrix}
=
\begin{pmatrix}
 4\Delta_p + c/2  &  6\Delta_p \\
 6\Delta_p & 4\Delta_p + 8\Delta_0{}^2
\end{pmatrix}.
\end{equation}

Next, we need to evaluate \begin{equation}
\vev{e^{2(Q-\alpha_1)\phi(\infty)}  e^{2\alpha_2\phi(1)} L_{-1}  e^{2\alpha\phi(0)} }.
\end{equation} To do this, we commute  $e^{2\alpha_2\phi(1)}$ and $L_{-1}$.
The term $L_{-1} e^{2\alpha_2\phi(1)}$ has $L_{-1}$ acting from the right on the primary $e^{2(Q-\alpha_1)\phi(\infty)}$, which annihilates it and gives zero. 
The computation of the commutator $[e^{2\alpha_2\phi(1)},L_{-1}]$ boils down to reinstating the position dependence by replacing $\phi(1)$ with $\phi(z)$, taking the derivative with respect to $z$, and setting $z=1$ again.  From the conformal invariance the $z$ dependence is just $z^{-h_{\alpha_1} + h_{\alpha_2}+h_{Q/2+ip}}$ where \begin{equation}
h_{\alpha}=\alpha(Q-\alpha),\qquad
\Delta_p=h_{Q/2+ip}=p^2 +(Q/2)^2.
\end{equation}
Then we have \begin{equation}
l_{\alpha_1,\alpha_2,\frac Q2+ip,\{1\}}=-h_{\alpha_1} + h_{\alpha_2}+\Delta_p, \qquad
r_{\{1\},\frac Q2-ip,\alpha_3,\alpha_4}=\Delta_p + h_{\alpha_3} - h_{\alpha_4}.
\end{equation} 
The next order terms are \begin{align}
l_{\alpha_1,\alpha_2,\frac Q2+ip,\{2\}}&=-h_{\alpha_1} + 2h_{\alpha_2}+\Delta_p, \\
r_{\{2\},\frac Q2-ip,\alpha_3,\alpha_4} &=\Delta_p + 2h_{\alpha_3} - h_{\alpha_4}, \\
l_{\alpha_1,\alpha_2,\frac Q2+ip,\{1,1\}}&=(-h_{\alpha_1} + h_{\alpha_2}+\Delta_p)(1-h_{\alpha_1} + h_{\alpha_2}+\Delta_p), \\
r_{\{1,1\},\frac Q2-ip,\alpha_3,\alpha_4} &=(\Delta_p + h_{\alpha_3} - h_{\alpha_4})(1+\Delta_p + h_{\alpha_3} - h_{\alpha_4}).
\end{align}

Combining the results,  we find \begin{equation}
F_{\alpha_1,\alpha_2,Q/2+ip,\alpha_3,\alpha_4}(q) 
= 1+ \frac{(-h_{\alpha_1} + h_{\alpha_2}+\Delta_p)(\Delta_p + h_{\alpha_3} - h_{\alpha_4})}{2\Delta_p}q+O(q^2).
\end{equation} The order $q^2$ term can be computed from the data shown above, but is too lengthy to be included here. 

It is tedious but not difficult to obtain terms of higher order in $q$ in the conformal block $F(q)$. 
Note that this is determined purely by the property of the Virasoro algebra, and the final result is expressed in terms of $h_{\alpha_i}=\alpha_i(Q-\alpha_i)$ and $\Delta_p$, that are the $L_0$ of the primary fields $e^{2\alpha_i\phi}$ and $e^{(Q+2ip)\phi}$, and the central charge $c$ only.   
It is instructive at this point to write a program in a computer algebra system of the reader's choice to compute the conformal block to the arbitrary order in $q$.

\section{A class of  four-dimensional theories}\label{sec:4}
In this section we introduce a class of four-dimensional \Nequals{2} supersymmetric gauge theories, commonly known as class S theories of type $\TASU(2)$ in the literature. This class of theories was first introduced in \cite{TAGaiotto:2009we}. For an extensive review, see e.g.~\cite{TATachikawa:2013kta}.
We start by quickly recalling the very basics of \Nequals{2} Lagrangians. 

\subsection{\Nequals{2} supersymmetric gauge theories}\label{sec:N2lag}
We assume the reader knows the basics of \Nequals{1} superfields. The \Nequals{2} theories we deal with can be obtained by imposing an $\TASU(2)_R$ symmetry that does \emph{not} commute with the \Nequals{1} supersymmetry apparent in the \Nequals{1} formalism.  

Let us start with the \Nequals{2} vector multiplet. This consists of an \Nequals{1} vector multiplet $V$ of a gauge group $G$, together with an \Nequals{1} chiral multiplet $\Phi$ in the adjoint representation of $G$. We consider the Lagrangian \begin{equation}
\frac{\Im\tau}{4\pi} \int d^4\theta \tr \Phi^\dagger e^{[V,\cdot]} \Phi +\int d^2\theta \frac{-\mathrm{i}}{8\pi}\tau \tr W_\alpha W^\alpha + cc.\label{vectorlag}
\end{equation} 
where $\tau=4\pi i/g^2 + \theta/2\pi$ is the complexified gauge coupling.
By expanding the superfields into components, we see that the gaugino $\lambda$ in $V$ and the chiralino $\psi$ in $\Phi$ have exactly the same couplings with the other fields, thus realizing $\TASU(2)_R$ symmetry. 

Next, we introduce the \Nequals{2} hypermultiplet, in the representation $R$ of the gauge group $G$. This consists of a pair of \Nequals{1} chiral multiplets $Q$, $\tilde Q$ in the representation $R$ and $\bar R$. The Lagrangian is \begin{equation}
\int d^4\theta (Q^\dagger{} e^V Q+ \tilde Q e^{-V} \tilde Q^\dagger{})
+(\int d^2 \theta \tilde Q \Phi Q + cc.) \label{hyperlag}
\end{equation} where $\mu$ is the mass term. Here, the $\TASU(2)_R$ symmetry rotates the scalar components of $Q$ and $\tilde Q^\dagger$. 

The Lagrangian above describes a massless hypermultiplet. To give a mass term, we can give a vev to $\Phi$ in the Lagrangian above. For example, take a pair of hypermultiplets $Q^a_i$ and $\tilde Q^i_a$ where $a=1,\ldots, N_c$ and $i=1,\ldots,N_f$.  This is in the bifundamental representation of $\TASU(N_c)\times \TAU(N_f)$, and as such we have the coupling \begin{equation}
\tilde Q^i_a \Phi^a_b Q^b_i + \tilde Q^i_a \underline{\Phi}_i^j Q^a_j 
\end{equation} in the Lagrangian, where $\Phi$ is in the adjoint of $\TASU(N_c)$ and $\underline{\Phi}$ is in the adjoint of $\TAU(N_f)$.  Now we regard $\underline{\Phi}$ and its associated \Nequals{1} vector multiplet as external, background fields and just give a vev $\vev{\underline{\Phi}}^i_j=m^i_j$.  
We end up having a mass term of the form $m^i_j \tilde Q^i Q_j.$ 
To preserve $\TASU(2)_R$ invariance, we require $[m,m^\dagger]=0$, which means that $m$ is diagonalizable.
It is known that this is the only way to give masses to hypermultiplets. 

When $m=0$, the $\TAU(N_f)$ symmetry is a global flavor symmetry. With generic nonzero diagonal $m$, this $\TAU(N_f)$ symmetry is further explicitly broken to $\TAU(1)^{N_f}$. We usually say that this mass term $m^i_j \tilde Q^i Q_j$ is associated to the flavor symmetry $\TAU(N_f)$. We often abuse the terminology and say that the theory has the $\TAU(N_f)$ flavor symmetry even when $m\neq 0$. 

Finally let us introduce the half-hypermultiplet. 
When the representation $R$ is pseudoreal, i.e.~when $R$ and $\bar R$ are equivalent as representations and when there is a gauge-invariant antisymmetric two-form $\epsilon_{ab}$, we can impose the condition \begin{equation}
Q_a = \epsilon_{ab} (\tilde Q^\dagger)^b
\end{equation} compatible with the $\TASU(2)_R$ invariance.  This is called a half-hypermultiplet in the representation $R$. Note that this consists of a single \Nequals{1} chiral multiplet in the representation $R$.

\subsection{Class S theories of type $\TASU(2)$}
\subsubsection{Construction from three-punctured spheres and cylinders}
In Sec.~\ref{sec:2}, we recalled the properties of two two-dimensional field theories. 
Very abstractly, two-dimensional field theories associate complex numbers to two-dimensional surfaces.
There, the essential point was to find the amplitude associated to a three-punctured sphere or a three-holed sphere, and the amplitude associated to a cylinder. 

Here, we introduce a way to associate four-dimensional field theories instead of complex numbers to two-dimensional surfaces. Again, the important point is to consider what to associate to a three-punctured sphere or a cylinder. 

\begin{figure}[h]
\[
\TAinc{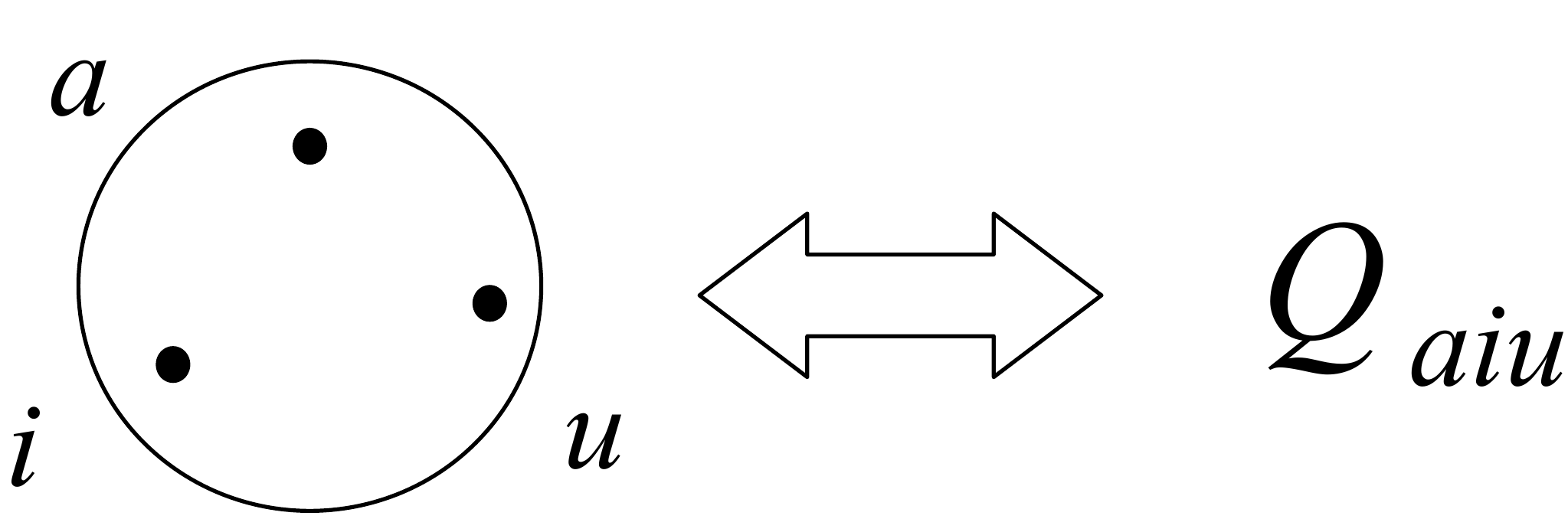}
\]
\caption{A three punctured sphere corresponds to $Q_{aiu}$, with $\TASU(2)^3$ symmetry.\label{fig:trifund}}
\end{figure}

First, take a three-punctured sphere, see Fig.~\ref{fig:trifund}. We associate an $\TASU(2)$ flavor symmetry for each of the three punctures.
The fundamental representation of $\TASU(2)$  is pseudoreal. Furthermore, the tensor product of an odd number of pseudoreal representations is pseudoreal.  Therefore, the \Nequals{1} chiral multiplet in the representation $(2,2,2)$ of $\TASU(2)_1\times \TASU(2)_2\times \TASU(2)_3$, which we denote as \begin{equation}
Q_{aiu},\qquad a=1,2; i=1,2; u=1,2
\end{equation} forms a half-hypermultiplet. Here $a$, $i$,  $u$ are the indices for $\TASU(2)_{1,2,3}$, respectively.

\begin{figure}[h]
\[
\TAinc{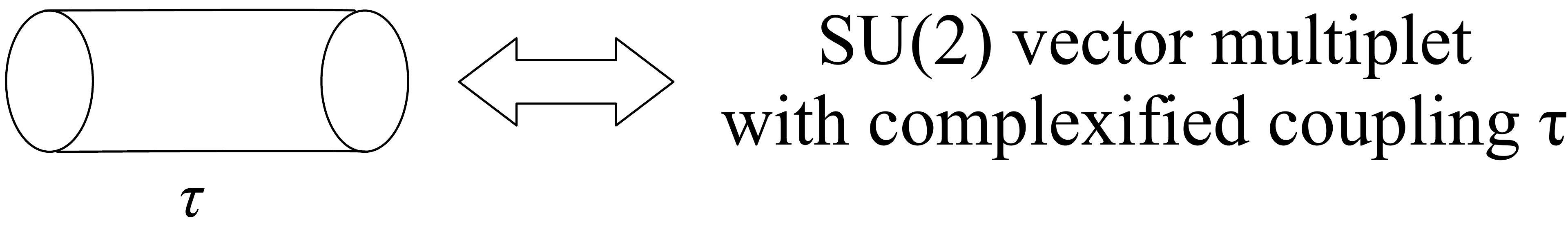}
\]
\caption{A cylinder with parameter $\tau$ corresponds to an $\TASU(2)$ vector multiplet.\label{fig:gauge}}
\end{figure}

Next, take a cylinder, see Fig.~\ref{fig:gauge}.  We assign to it a complex number $\tau=4\pi i/g^2 + \theta/2\pi$, where $g$ is real and $\theta\sim \theta+2\pi$.  We then associate to it an \Nequals{2} vector multiplet with gauge group $\TASU(2)$, whose complexified coupling is $\tau$. 

Now, given a collection of three-punctured spheres, we pick two punctures. Let us say that the first puncture is at $z=0$ in a local coordinate $z$ and the second is at $w=0$ in a local coordinate $w$. 
We take a cylinder with parameter $\tau$, and make the identification \begin{equation}
z w= e^{2\pi i \tau}.
\end{equation}  We also often use the notation $q=e^{2\pi i\tau}$.

Correspondingly, we perform the following operation on the four-dimensional gauge theory side.  By choosing two punctures, we picked up two $\TASU(2)$ flavor symmetries. We now couple them to a dynamical $\TASU(2)$ gauge field, whose complexified coupling constant is $\tau$.

\subsubsection{One-punctured torus}
\begin{figure}[h]
\[
\TAinc{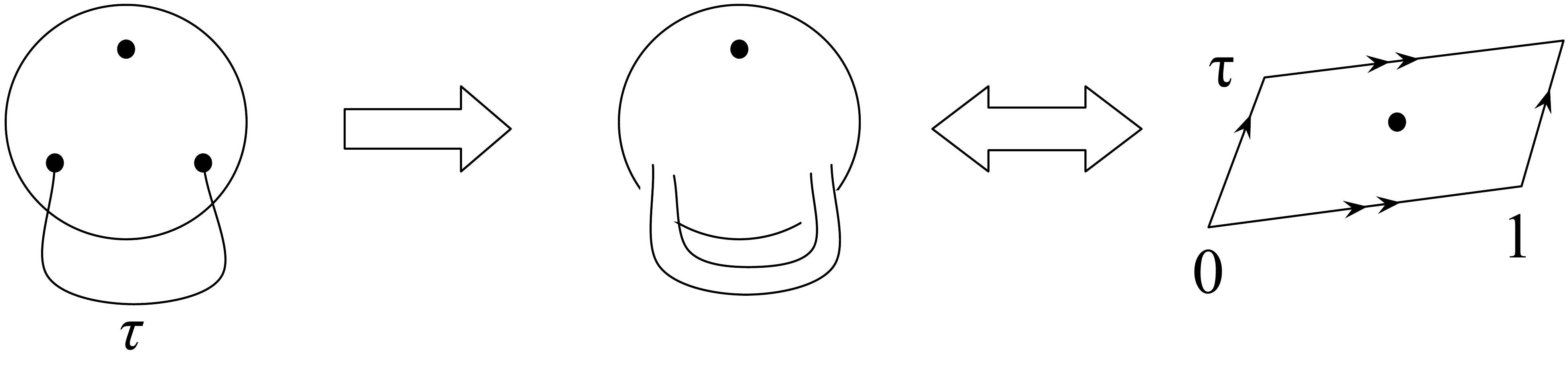}
\]
\caption{A punctured torus can be obtained from gluing two points of a three-punctured sphere.\label{fig:torus}}
\end{figure}

Take one three-punctured sphere, and connect two punctures out of three, by a cylinder of parameter $\tau$.  
The result is a torus of modulus $\tau$ with a puncture, see Fig.~\ref{fig:torus}. 
As a gauge theory operation, we start from a trifundamental $Q_{aiu}$. Pick the indices $a$ and $i$, and we couple it to a single $\TASU(2)$ gauge field. 
We now regard two $\TASU(2)$ symmetries acting on $a$ and $i$ as one and the same; as a doublet times a  doublet is a triplet plus a singlet, we relabel $Q_{aiu}$ as $A_{Iu}$ ($I=1,2,3$) and $H_{u}$, where the index $u=1,2$ is still for an $\TASU(2)$ flavor symmetry.  

We end up with an \Nequals{2} $\TASU(2)$ gauge theory with a triplet hypermultiplet formed by $A_{I}:=A_{I,u=1}$, $\tilde A_I:=A_{I,u=2}$ and a decoupled hypermultiplet formed by $H:=H_{u=1}$, $\tilde H:=H_{u=2}$.  An \Nequals{2} gauge theory with gauge group $G$, together with a hypermultiplet in the adjoint representation of $G$, has an enhanced \Nequals{4} supersymmetry, and commonly known just as \Nequals{4} supersymmetric Yang-Mills. We can give a mass term using $\TASU(2)$ flavor symmetry, still preserving \Nequals{2} supersymmetry. The resulting theory is often called \Nequals{2^*} theory. 

Note that the gauge theory we obtained has zero one-loop beta function. The high-degree of supersymmetry guarantees that the beta function is zero  even non-perturbatively. Therefore the complexified coupling $\tau$ remains a dimensionless parameter in the gauge theory.  

An important property of \Nequals{4} super Yang-Mills is its S-duality: the theory with gauge group $G$ with coupling constant $\tau$ is equivalent to the theory with dual gauge group $G^\vee$ with coupling constant $-1/\tau$.  We only use the case $G=\TASU(2)$, where the dual gauge group $G^\vee$ happens to be the same as the original one, so $G=G^\vee=\TASU(2)$. 

This has a nice interpretation in our geometric construction of gauge theories: we can construct a one-punctured torus with modular parameter $\tau$ in two ways from a three-punctured sphere and a cylinder, namely with a cylinder with parameter $\tau$ or with parameter $-1/\tau$.

Correspondingly, for this once-punctured torus, we have two gauge theories associated under our rule: \Nequals{4} theory with gauge group $\TASU(2)$, with complexified gauge coupling $\tau$ or the same theory with gauge coupling $-1/\tau$.  The S-duality guarantees that these two theories are the same. 

\subsubsection{Four-punctured sphere}
Next, take two three-punctured spheres, and connect one puncture from a sphere and another puncture from another sphere, with a cylinder with parameter $\tau$.  This gives a four-punctured sphere. 

In the gauge theory language, we have two trifundamentals $Q_{aiu}$ and $Q'_{usx}$. The indices $a,i,s,x=1,2$ are for four $\TASU(2)$ symmetries, and we have a dynamical $\TASU(2)$ gauge multiplet acting on the index $u=1,2$.  
The one-loop beta function is zero. Again, the beta function is zero even non-perturbatively, and the complexified coupling constant is a genuine dimensionless parameter of the theory. 

We can reorganize the chiral matter fields into $q_{uI}$ where $u=1,2$ and $I=1,\ldots, 8$. The superpotential coupling is \begin{equation}
\delta^{IJ} q_{uI} \Phi_{uv} q_{vJ}
\end{equation}  where $\Phi$ is the adjoint chiral scalar of the $\TASU(2)$ gauge multiplet. The indices $uv$ are symmetric, and therefore $\delta_{IJ}$ is symmetric too. This means that the flavor symmetry is $\TASO(8)$. 
This theory is often called $\TASU(2)$ gauge theory with $N_f=4$ flavors. 
The four $\TASU(2)$s are subgroups of this $\TASO(8)$: \begin{equation}
\TASU(2)_a\times \TASU(2)_i \times  \TASU(2)_s \times \TASU(2)_x \subset \TASO(8)
\end{equation} and the chiral matter fields transform as \begin{equation}
2_a \otimes 2_i \oplus 2_s\otimes 2_x = 8_V
\end{equation} where $8_V$ is the vector representation of the $\TASO(8)$ flavor symmetry. 

\begin{figure}[h]
\[
\TAinc{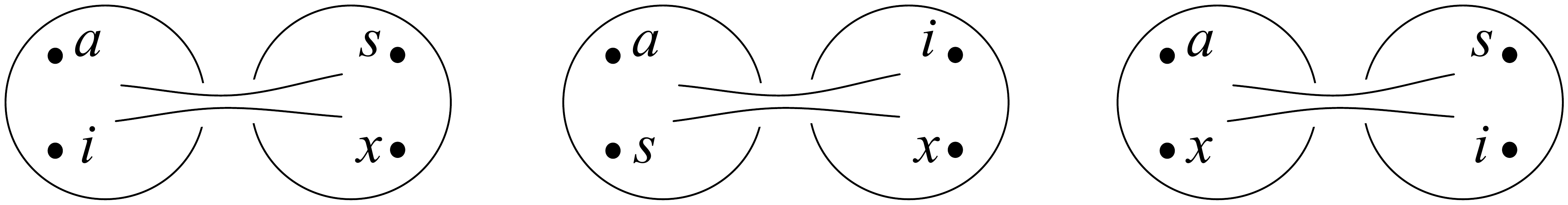}
\]
\caption{A four-punctured sphere  can be decomposed in three different ways.\label{fig:triality}}
\end{figure}

Now, a four-punctured sphere can be obtained in three distinct ways from two three-punctured spheres, see Fig.~\ref{fig:triality}. We described one already. Two others also give $\TASU(2)$ theories with $N_f=4$ flavors, but with different $\TASO(8)$ representation and with different coupling constants.  Namely, the matter fields in \begin{equation}
2_a \otimes 2_s \oplus 2_i\otimes 2_x = 8_S
\end{equation} with coupling $q'=1/q$ 
and the matter fields in \begin{equation}
2_a \otimes 2_x \oplus 2_s\otimes 2_i = 8_C
\end{equation} with coupling $q''=1-q$. 
Here, $8_S$ and $8_C$ are positive and negative chirality spinors of $\TASO(8)$, and
we used the exponentiated complexified coupling constants $q=e^{2\pi i \tau}$, 
$q'=e^{2\pi i \tau'}$, 
$q''=e^{2\pi i \tau''}$.
Again, the S-duality guaranteeing the equivalence of these three descriptions has been known for quite some time. 

\subsubsection{General consideration}
Let us recapitulate what we have introduced so far. 
First, we gave a method to construct an \Nequals{2} gauge theory. The data was encoded in terms of the three-punctured spheres and cylinders connecting pairs of punctures.  As can be easily checked, every gauge group corresponding to any cylinder has zero one-loop beta function. The \Nequals{2} supersymmetry then guarantees that all beta functions are zero even non-perturbatively, and therefore all gauge coupling constants remain genuine dimensionless parameters of the theory.   The punctures that remain unused for connection via cylinders provide $\TASU(2)$ flavor symmetries. With $k$ punctures we have $\TASU(2)^k$ flavor symmetries, and if desired, we can turn on the corresponding $k$ mass parameters.  At this point, we have a Lagrangian field theory for a given Riemann surface with punctures, with the decomposition into three-punctured spheres specified. 

Second, we considered a torus with one puncture can be constructed from two three-punctured spheres in two ways.  The corresponding gauge theories were both \Nequals{4} super Yang-Mills with gauge group $\TASU(2)$ with a decoupled hypermultiplet, but with different coupling $\tau$ and $-1/\tau$.  They are known to be S-dual. Therefore, a torus with one puncture in fact corresponded to a single theory.

Third, we considered a sphere with four-punctures. This can be constructed in three ways from two three-punctured spheres. The three corresponding gauge theories were $\TASU(2)$ theories with $N_f=4$ flavors, but with hypermultiplets in distinct $\TASO(8)$ flavor representations and with distinct coupling constants. Again, they are known to be S-dual. Therefore, a four-punctured sphere in fact corresponded to a single theory.

In general, given a single Riemann surface with a number of punctures, there are multiple ways to cut it into three-punctured spheres and cylinders.  Each of such decompositions gives rise to a distinct Lagrangian of \Nequals{2} supersymmetric gauge theory. 
However, by combining the two S-dualities recalled above, any such decompositions can be related. Therefore, we conclude that a single Riemann surface with a number of punctures in fact corresponds to a single quantum field theory, of which various distinct Lagrangians are just avatars.

\section{Two supersymmetric backgrounds}\label{sec:part}
In this section we first discuss the general idea of supersymmetric localization, following the approach pioneered by \cite{TAFestuccia:2011ws} and reviewed in \volcite{DU}. We then discuss two explicit supersymmetric backgrounds, one based on $S^1\times S^3$ and another based on $S^4$.  More details on the former and the latter can be found in \volcite{RR} and in \volcite{HO},  respectively.

\subsection{General yoga of supersymmetric localization}
The energy-momentum tensor $T_{\mu\nu}$ of a quantum field theory describes how it couples to an external metric perturbation. Namely, let $Z$ be the partition function as a functional of the metric. Then we have  
$\delta \log Z=\int \delta g_{\mu\nu} T^{\mu\nu} d^dx$.
By integrating this small variation, we know how a quantum field theory behaves in a general curved manifold.  

In a supersymmetric theory, the energy-momentum tensor $T_{\mu\nu}$ sits in a supermultiplet, containing the supercurrent $S_{\mu\alpha}$ and other components, depending on the dimension of the spacetime, the number of supersymmetries, and other subtler properties.  In this review we are interested in \Nequals{2} theory in  four dimensions.  We then have the R-current $J_{\mu}^R$ and a scalar component $X$, both in the adjoint of $\TASU(2)_R$. 

The energy-momentum tensor $T_{\mu\nu}$ knows how the theory couples to the metric $g_{\mu\nu}$. Similarly, the supercurrent $S_{\mu\alpha}$ knows how it couples to the gravitino background $\psi_{\mu\alpha}$. The R-current $J_{\mu}^R$ knows how it couples to the R-symmetry background $A_\mu^R$, and the scalar component $X$ knows how it couples to the scalar background $M$. 
In short, the supermultiplet $(T_{\mu\nu}, S_{\mu\alpha}, J_{\mu}^R, X, \ldots)$ knows how it couples to the external supergravity background $(g_{\mu\nu}, \psi_{\mu\alpha}, A_{\mu}^R, M,\ldots)$.  
There are many definitions of supermanifolds in the mathematical literature, but from the point of view of the supersymmetric field theories, the most natural super-version of a curved manifold is a manifold with the full supergravity background specified. 
This point of view was emphasized first in \cite{TAFestuccia:2011ws} and reviewed in more detail in  \volcite{DU}.

Similarly, if a theory has a global flavor symmetry group $G$, it has a flavor current $J_\mu$, which knows how the theory couples to the flavor symmetry background $A_\mu$. When the theory is \Nequals{2} supersymmetric in four dimensions, $J_\mu$ sits in a supermultiplet containing a scalar operator $K$, that knows how to couple to a background scalar $\Phi$, where both $K$ and $\Phi$ are adjoint of $G$. In fact, the mass term of \Nequals{2} theory is just a special case of this construction, as we recalled in Sec.~\ref{sec:N2lag}.

Now, given a quantum field theory on a curved manifold with isometry $\xi$,  we have $\vev{\delta_\xi O}=0$ for any operator $O$. Similarly, given a supersymmetric field theory on  a  supersymmetric background, i.e.~a supergravity background with at least one superisometry $\epsilon$, we have $\vev{\delta_\epsilon O}=0$ for any operator $O$.  

It often happens that many natural bosonic operators in the Lagrangian can be written as $\delta_\epsilon O$. 
For example, suppose a coupling $\lambda$ in the Lagrangian multiplies an operator $X=\delta_\epsilon O$.
Then, the partition function is independent of $\lambda$, because \begin{equation}
\frac{\partial}{\partial \lambda}\log Z=\vev{X}=\vev{\delta_\epsilon O}=0.
\end{equation}
Also, note that the superisometry variation of the supercurrent itself, $\delta_\epsilon S_{\mu \alpha}$, is given by a linear combination of the energy-momentum tensor and other bosonic components of the supermultiplet. This means that a certain variation of the metric can be combined with corresponding particular variations of R-symmetry and scalar backgrounds so that the partition function is independent of it. This makes the partition function on a supersymmetric background oblivious to detailed choice of the metric.  Sometimes it depends only on the topology or the complex structure of the spacetime. When $\delta_\epsilon^2$ generates a bosonic isometry of the background, the partition function only depends on the topological property of that isometry, etc. 

It also often happens that we can choose a fermionic operator $O$ such that \begin{equation}
\delta_\epsilon{}^2O=0, \qquad
\delta_\epsilon O \simeq \sum_{\psi} |\delta\psi|^2.
\end{equation} where $\psi$ runs over the dynamical fermion fields in the theory. 
We then consider the deformation \begin{equation}
S\to S(t)=S+t\int d^dx \delta_\epsilon O.
\end{equation} Thanks to $\delta_\epsilon{}^2O=0$, the deformed Lagrangian is still invariant under the superisometry $\epsilon$. 
Then \begin{equation}
\frac{\partial}{\partial t}\log Z(t)=\int d^dx\vev{\delta_\epsilon O} =0
\end{equation} for arbitrary $t$. 
In the large $t$ limit, the integral localizes to the configurations satisfying \begin{equation}
\delta\psi=0\label{BPSeq}
\end{equation} and fluctuations around them.  For this reason we often call $\delta_\epsilon O$ as the localizing term. 

We parameterize the solutions to \eqref{BPSeq}  by a space $\CalM=\sqcup_i \CalM_i$ where $i$ is some label distinguishing the components.
Then we have the equality \begin{equation}
Z=\sum_i \int_{\CalM_i} Z_\text{classical} Z_\text{quadr. fluct.},
\end{equation}
where $Z_\text{classical}$ is the exponential of the classical action evaluated at a configuration satisfying \eqref{BPSeq}, and $Z_\text{quadr. fluct.}$ is the result of the Gaussian integrals of bosonic and fermionic fluctuations.  The interaction terms do not contribute in the $t\to \infty$ limit.  

Often, each of the components $\CalM_i$ is finite dimensional, and therefore the partition function $Z$ is given as a sum of explicit multiple integrals. Evaluating it is still a formidable task, but is infinitely simpler than the original infinite-dimensional path integral expression. 

Now, let us discuss two particular classes of supersymmetric backgrounds for four-dimensional \Nequals{2} theories, on which the localization has been worked out in detail. The first is $S^1\times S^3$ and the second is $S^4$. We do not discuss the detailed derivations of the facts mentioned below. Happily, all the details can be found in \volcite{RR} for $S^1\times S^3$ and in \volcite{HO} for $S^4$.

\subsection{$S^1\times S^3$}\label{sec:part-sci}
First, we consider \Nequals{2} superconformal theory on $S^1\times S^3$. 
An \Nequals{2} superconformal theory has $\TASU(2)_R\times \TAU(1)_R$ R-symmetry.
A class of supersymmetric backgrounds, where the supersymmetric localization can be performed, is specified by the ratio $\beta$  of the radii of $S^1$ and $S^3$, and two holonomies of $\TASU(2)_R$ and $\TAU(1)_R$ around $S^1$. 
In total there are three parameters, commonly denoted by $(p,q,t)$. Note that this $q$ is independent of the exponentiated complexified coupling, also often denoted by $q$ and is equal to $e^{2\pi i\tau}$. 
In fact, the vector multiplet Lagrangian is of the form $\sim \tau\delta_\epsilon O$, and therefore the partition function on this background is independent of the complexified gauge coupling $\tau$. 

For brevity of the exposition, we only use the one-dimensional slice where $p=0$, $q=t$.  In this case $q=e^{-\beta}$.  
When the \Nequals{2} theory in question has a flavor symmetry $G$, we can choose an element $g\in G$ such that there is a background $G$ gauge field around $S^1$ given by $g$.   Then the partition function on this background is just a function of $q=e^{-\beta}$ and $g\in G$. 

By considering the $S^1$ direction as the time direction, the partition function can be written as \begin{equation}
Z(q,g)= \tr_{\CalH} (-1)^F q^{\Delta-R} g.
\end{equation} Here, $\CalH$ is the Hilbert space of the superconformal theory on $S^3$, which is equivalent to the space of point-like operators of the conformal theory on the flat space, via the state-operator correspondence.  $\Delta$ is then the scaling dimension of the operator, and $R$ is the third component of the $\TASU(2)_R$ charge.  From this structure this partition function is often called the superconformal index. For four-dimensional quantum field theories, this concept was introduced in \cite{TARomelsberger:2005eg,TAKinney:2005ej}.

Correspondingly, the computation of $Z(q,g)$ for an \Nequals{2} Lagrangian field theory can be done either by counting the operators on a flat space, or by performing the path integral on $S^1\times S^3$. In most of th  literature it is done using the operator approach; for a through discussion for \Nequals{2} case, see \cite{TAGadde:2011uv}. For a path-integral approach for \Nequals{4}, see \cite{TANawata:2011un}.

Either way, we find the following results.  We first represent  the holonomy $g\in G$ around $S^1$ using complex numbers in the form \begin{equation}
g=(z_1,\ldots, z_r) \in \TAU(1)^r \subset G
\end{equation} where $r$ is the rank of $G$.  We can have an arbitrary flavor holonomy. The gauge holonomy parameterizes the BPS configurations over which we integrate.

For an \Nequals{2} hypermultiplet consisting of \Nequals{1} chiral multiplet in a representation $R$ of a symmetry $G$, the partition function is \begin{equation}
Z_R(q,g)=\prod_{n\ge 0}\prod_{w}\frac{1}{1-q^{n+1/2}z^w}\label{sci-hyper}
\end{equation} where $w=(w_1,\ldots, w_r)$ runs over weights of $R$ and $z^w:=\prod z_i{}^{w_i}$.

For a gauge theory with gauge group $G$, the partition function is  \begin{equation}
Z(q)=\frac{1}{|W_G|} \oint \prod_{i=1}^r \frac{dz_i}{2\pi\sqrt{-1}z_i}
\prod_\alpha (1- z^\alpha)  K(z)^{-2}  \times \text{(matter contribution)}\label{sci-gauge}
\end{equation} where \begin{equation}
K(z)^{-1}=\prod_{n\ge 0} \left[(1-q^{n+1})^r \prod_\alpha (1-q^{n+1} z^\alpha)\right].\label{sci-K}
\end{equation}
Here, the product on $\alpha$ runs over the roots $\alpha$ of $G$, $|W_G|$ is the order of the Weyl group of $G$, and the integral takes the residues at the origin. 

As an example, the partition function of \Nequals{4} supersymmetric Yang-Mills of gauge group $\TASU(N)$, considered as an \Nequals{2} theory with $\TASU(2)$ flavor symmetry, is given as follows. The $\TASU(2)$ element $g$ is written as $(a,a^{-1})\in \TASU(2)$. 
The $\TASU(N)$ element is written as $(z_1,\ldots, z_N)\in \TASU(N)$, with the understanding that $\prod z_i=1$. 
Then, we have \begin{multline}
Z(q,a)=\frac1{N!} \oint \prod_{i=1}^{N-1} \frac{dz_i}{2\pi\sqrt{-1}z_i}
\prod_{i,j}(1-z_i/z_j)^2\times \\
\frac{\prod_{n\ge 0} (1-q^{n+1})^{2r} \prod_{i,j}(1-q^{n+1} z_i /z_j)^2}
{\prod_{\pm} \prod_{n\ge 0} (1-q^{n+1/2} a^{\pm1})^{r} 
\prod_{i,j}(1-q^{n+1/2} z_i /z_j a^{\pm1}) }.
\end{multline}
This looks complicated, but it is quite explicit, and its $q$-expansion can be readily computed. 

\subsection{$S^4$}\label{sec:S4}
As a second supersymmetric background, we consider $S^4$. The localization on a round $S^4$ was first done in \cite{TAPestun:2007rz}, and this was later extended to squashed $S^4$ in \cite{TAHama:2012bg}. We call both backgrounds just $S^4$. A review of the localization on general backgrounds with two isometries can be found in  \cite{TAPestunLocalizationReview}.

\subsubsection{General structure}
To describe the localization, we first write $S^4$ as a hypersurface in a five-dimensional space parameterized by a real number $x$
and two complex numbers $z_{1,2}$: \begin{equation}
x^2 + |z_1|^2 + |z_2|^2 = 1.
\end{equation} Write $z_i = r_i e^{\sqrt{-1} \phi_i}$, and 
we endow $S^4$ with background supergravity fields invariant under the arbitrary shift of $\phi_1$ and $\phi_1$. 
The metric and other supergravity fields can be suitably chosen so that 
there is a superisometry $\epsilon$ such that $\delta_{\epsilon}{}^2$ generates a rotation \begin{equation}
\phi_i \to \phi_i + \epsilon_i.
\end{equation} Note that $x=1$, $z_i=0$ and $x=-1$, $z_i=0$ are two fixed points under this rotation.
We call them the north pole and the south pole. 

The field configurations that contribute to the partition function under the localization are as follows. 
For the hypermultiplets, the vev should be all zero.
For the vector multiplets, the adjoint scalar $\Phi$ is such that $\Phi$  can have a non-zero spacetime-independent vev but $\bar\Phi=0$. 
This vev is customarily denoted by $a$. We take the convention that $a$ is purely imaginary.
When the vector multiplet is non-dynamical, this vev $a$ gives the mass term of the hypermultiplets, and is often denoted by $m$. 
In addition, there can be point-like instantons supported on the north pole and point-like anti-instantons supported on the south pole. 

Correspondingly, the partition function of the theory on $S^4$ with gauge group $G$ has the form \begin{equation}
Z=\frac{1}{|W_G|}  \int d^r a Z(a) \overline{Z(a)}\label{pestun-integral}
\end{equation} where $|W_G|$ is the order of the Weyl group, the integral is over the space of vevs of the real part of $\Phi$ that is gauge-fixed to lie on the Cartan subalgebra. Then $Z(a)$ is the contribution from the northern hemisphere given the vev $a$, and its complex conjugate $\overline{Z(a)}$ gives the contribution from the southern hemisphere. 

The contribution $Z(a)$ is often called Nekrasov's partition function, and is composed of the following ingredients: \begin{equation}
Z(a)=Z_\text{classical}(a) Z_\text{one-loop}(a) Z_\text{instanton}(a).
\end{equation}
In the following, we give explicit forms of these three factors. We will be cavalier about the overall factors independent of the vev of the adjoint scalars, mass parameters and gauge coupling constants. 

\subsubsection{Classical and one-loop factors}
The first factor is the exponentiated classical action, given by the product of \begin{equation}
e^{-\frac{1}{\epsilon_1\epsilon_2} 2\pi \sqrt{-1} \tau \vev{a,a}}
\end{equation} over various gauge multiplets, where $\tau$ is the complexified gauge coupling. 
The second factor is the one-loop contributions from the vector multiplets and the hypermultiplets.
From a gauge multiplet, we have 
\begin{equation}
Z_{\text{one-loop}}^G=\prod_{\alpha>0} \frac{1}
{\Gamma_2(\alpha\cdot a + \epsilon_1+\epsilon_2|\epsilon_1,\epsilon_2)
\Gamma_2(\alpha\cdot a|\epsilon_1,\epsilon_2)} 
\end{equation} where the product runs over the positive roots $\alpha$.
From a hypermultiplet in the representation $R$, we have \begin{equation}
Z_{\text{one-loop}}^R=\prod_{w} 
\Gamma_2( w\cdot a + \frac{\epsilon_1+\epsilon_2}2|\epsilon_1,\epsilon_2)\label{nek-hyper}
\end{equation} where the product runs over the weights of $R$. For a half-hypermultiplet in $R$, the product runs over arbitrary half of the weights of $R$; the final contribution to the instanton integral \eqref{pestun-integral} does not depend on this split into halves. 

The mass term of hypermultiplets can be incorporated by giving a vev to the scalar in the flavor background gauge multiplet. Effectively, this just replaces $w\cdot a \to w\cdot a + m$. 
Note also that in our convention $a$ is purely imaginary but  $\epsilon_{1,2}$ are  real. Therefore, to get the contribution $\overline{Z(a)}$ from the southern hemisphere, we just replace $a$ by $-a$ in the argument of the double Gamma function.

\subsubsection{Instanton contributions}
The instanton contribution is much more complicated to present and the explicit form is only known for $\TASU$ gauge groups and with full hypermultiplets, due to various technical problems. 
Even for $\TASU$ gauge groups, the computation involves a certain regularization that introduces spurious contributions, which is often said to come from replacing $\TASU$ groups by $\TAU$ groups and therefore has the name $\TAU(1)$ factors.  Here we just quote the known results, including the spurious or $\TAU(1)$ factor.
For a more detailed discussion, see \cite{TATachikawaInstantonReview}.

A point-like instanton configuration of a $\TAU(N)$ gauge multiplet is labeled by an $N$-tuple of Young diagrams $\vec Y=(Y_1,\ldots, Y_N)$. A Young diagram $Y$ is just a non-decreasing sequence of natural numbers $Y=(\lambda_1 \ge \lambda_2 \ge \cdots)$. 
Here we visualize them by considering $\lambda_i$ as the height of the $i$-th column from the left, and
we set $\lambda_i=0$ when $i$ is larger than the width of the diagram. 
This is not the standard convention in the literature of Young diagrams, but is the standard one in the instanton computations. 
We denote by $|Y|$ the number of boxes, i.e.~$|Y|=\sum \lambda_i$
and then we define $|\vec Y|=\sum |Y_i|$. 

We denote by $Y^T$ its transpose, $Y^T=(\lambda_1'\ge \lambda_2' \ge \cdots)$.
For a box $s$ at the coordinate $(i,j)$,  its arm-length $A_Y(s)$ and the leg-length $L_Y(s)$ are defined to be \begin{equation}
A_Y(s)=\lambda_i-j, \qquad L_Y(s)=\lambda_j'-i.
\end{equation} Note that $s$ can be outside of the Young diagram $Y$. We then let \begin{equation}
E(a,Y_1,Y_2,s)=a-\epsilon_1 L_{Y_2}(s)+\epsilon_2(A_{Y_1}(s)+1).
\end{equation}

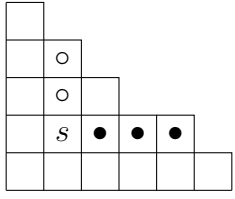
\begin{figure}
\centering
\begin{tikzpicture}[scale=0.25]
\draw (0,0)--(0,10);
\draw (2,0)--(2,10)--(0,10);
\draw (4,0)--(4,8)--(0,8);
\draw (6,0)--(6,6)--(0,6);
\draw (8,0)--(8,4)--(0,4);
\draw (10,0)--(10,4)--(8,4);
\draw (12,0)--(12,2)--(0,2);
\draw (0,0)--(12,0);
\node (A) at (3,3) {$s$};
\foreach\i in {5,7,9} 
{
	\node at (\i,3) {$\bullet$};
}
\foreach\i in {5,7} 
{
	\node at (3,\i) {$\circ$};
}
\end{tikzpicture}
\caption{Young diagram $Y=(5,4,3,2,2,1)$ and a box $s=(2,2)$, and its arm and leg length.}
\end{figure}

Now we can finally write down the instanton contribution. For example, consider an $\TASU(N)\times \TASU(M)$ gauge theory with a bifundamental hypermultiplet. Its instanton contribution is a sum over possible point-like instanton configurations:\begin{equation}
Z_\text{instanton}(\vec a;\vec b)=\sum_{\vec Y,\vec W} q^{|Y|} q'{}^{|W|} Z(\vec a,\vec Y;\vec b,\vec W)
\end{equation} where $\vec a=(a_1,\ldots,a_N)$ and $\vec b=(b_1,\ldots, b_M)$ are the vevs of the real part of the adjoint scalars in the vector multiplets of $\TASU(N)$ and $\TASU(M)$,
and $\vec Y=(Y_1,\ldots,Y_N)$, $\vec W=(W_1,\ldots, W_M)$ label the point-like instanton configurations. 
The prefactors $q$, $q'$ are given by $q=e^{2\pi \sqrt{-1} \tau}$
and $q'=e^{2\pi \sqrt{-1} \tau'}$ where $\tau$, $\tau'$ are the complexified gauge couplings of two gauge factors.

The contribution from each fixed point, $Z(\vec a,\vec Y;\vec b,\vec W)$, is given by the product of the contributions from each multiplet. The vector multiplet of $\TASU(N)$ contributes by \begin{equation}
Z_\text{instanton}^\text{vector,$\TASU(N)$}(\vec a,\vec Y)=\frac{1}{
\prod_{i,j=1}^N
\prod_{s\in Y_i}E(a_i-a_j,Y_i,Y_j,s)
\prod_{t\in Y_j}(\epsilon_1+\epsilon_2  -  E(a_j-a_i,Y_j,Y_i,t))
},
\end{equation} and the contribution from $\TASU(M)$ can be obtained similarly. The contribution from the bifundamental of mass $m$ is \begin{multline}
Z_\text{instanton}^\text{hyper,bifundamental}(\vec a,\vec Y;\vec b,\vec W;m)=\\
\prod_{i}^N \prod_j^M
\prod_{s\in Y_i}( E(a_i-b_j,Y_i,W_j,s)- m-\frac{\epsilon_1+\epsilon_2}2)
\prod_{t\in W_j}( E(b_j-a_i,W_j,Y_i,t)-m+\frac{\epsilon_1+\epsilon_2}2).\label{nek-bifund}
\end{multline}
The contributions from an adjoint hypermultiplet of $\TASU(N)$ can be obtained by setting $\vec a=\vec b$ 
and $\vec Y=\vec W$. Similarly, the contributions from $N_f$ fundamental hypermultiplet of $\TASU(N)$ can be obtained by letting $M=N_f$ and regarding the $\TASU(M)$ part as a background gauge field, setting $W_i=0$. Then $\vec b$ becomes the $\TASU(N_f)$ part of the mass parameters. 
As an example, the partition function of $\TASU(N)$ gauge theory with an adjoint hypermultiplet of mass $m$ is \begin{multline}
Z(\tau,m)=\frac{1}{N!}\int d^{N-1} a |q^{-\frac{1}{\epsilon_1\epsilon_2}\sum a_i^2}|^2 \\
\times \prod_{i,j} \frac
{\Gamma_2((a_i-a_j) + m+\frac{\epsilon_1+\epsilon_2}2|\epsilon_1,\epsilon_2)
\Gamma_2((a_i-a_j)-m+\frac{\epsilon_1+\epsilon_2}2|\epsilon_1,\epsilon_2)}
{\Gamma_2((a_i-a_j) + \epsilon_1+\epsilon_2|\epsilon_1,\epsilon_2)
\Gamma_2((a_i-a_j)|\epsilon_1,\epsilon_2)} \\
\times \left| \sum_{\vec Y} q^{|Y|}
\prod_{i,j} \frac
{
\prod_{s\in Y_i}( E(a_i-a_j,Y_i,Y_j,s)-m-\frac{\epsilon_1+\epsilon_2}2)
\prod_{t\in Y_j}( E(a_j-a_i,Y_j,Y_i,t)- m+\frac{\epsilon_1+\epsilon_2}2)
}
{
\prod_{s\in Y_i}E(a_i-a_j,Y_i,Y_j,s)
\prod_{t\in Y_j}(\epsilon_1+\epsilon_2  -  E(a_j-a_i,Y_j,Y_i,t))
}
\right|^2
\end{multline}

This expression is horrible, but quite explicit nonetheless. The matter content of the theory is the same as \Nequals{4} super Yang-Mills with gauge group $\TASU(N)$, and the supersymmetry on this supergravity background is believed to enhance when $m=\pm (\epsilon_1-\epsilon_2)/2$. 
Then the expression simplifies greatly \cite{TAOkuda:2010ke}.
In this case, the instanton correction is trivial, because when $|Y|\ge 1$, there is a box that causes at least one factor of the numerator to be zero. For each choice of $(i,j)$, the one-loop factor gives \begin{equation}
\frac{
\Gamma_2((a_i-a_j)+\epsilon_1|\epsilon_1,\epsilon_2)
\Gamma_2((a_i-a_j)+\epsilon_2|\epsilon_1,\epsilon_2)
}
{
\Gamma_2((a_i-a_j)|\epsilon_1,\epsilon_2)
\Gamma_2((a_i-a_j)+\epsilon_1+\epsilon_2|\epsilon_1,\epsilon_2)
}
=a_i-a_j.
\end{equation} Then the integral boils down to \begin{equation}
Z(\tau,\frac{\epsilon_1-\epsilon_2}2)=\frac{1}{N!}\int d^{N-1} a \prod_{i,j}(a_i-a_j) |q^{-\frac{1}{\epsilon_1\epsilon_2}\sum a_i^2}|^2 
=\int d^{N^2-1} A |q|^{-\frac{2}{\epsilon_1\epsilon_2} \tr AA^\dagger},
\end{equation} that is, this is just an Hermitean random matrix model.  This  was originally observed in \cite{TAErickson:2000af,TADrukker:2000rr}, and the $S^4$ localization of \cite{TAPestun:2007rz} was originally conceived to derive this fact quantum field theoretically.

\section{Two correspondences and an interpretation}\label{sec:correspondences}
Here comes the crux: we compute the partition functions of the class of 4d \Nequals{2} theories
associated to Riemann surfaces with punctures
introduced in Sec.~\ref{sec:4} on two supersymmetric backgrounds introduced in Sec.~\ref{sec:part}.
We will find that they are given by the two two-dimensional theories introduced in Sec.~\ref{sec:2}. 

\subsection{$S^1\times S^3$ and the two-dimensional Yang-Mills}\label{sec:2d4d-sci}
Let us first consider the partition function on $S^1\times S^3$.
Take a three-punctured sphere. In Sec.~\ref{sec:4} we associated to it a half-hypermultiplet in $2\otimes 2\otimes 2$ of $\TASU(2)_1\times \TASU(2)_2\times \TASU(2)_3$. 
Its partition function $Z(a_1,a_2,a_3)$ on $S^1\times S^3$, when the holonomy of $\TASU(2)_i$ is $(a_i,1/a_i)\in \TASU(2)_i$, is given using \eqref{sci-hyper} by \begin{equation}
Z(a_1,a_2,a_3)=\prod_{\pm\pm\pm}\prod_{n\ge 0} \frac1{1-q^{n+1/2} a_1^{\pm1}a_2^{\pm1}a_3^{\pm1} }.\label{sci-trifund-prod}
\end{equation}  It so happens that it has an alternative expression as an infinite sum \begin{equation}
Z(a_1,a_2,a_3)=\frac{K(a_1)K(a_2)K(a_3)}{K_0} \sum_{n\ge 0} \frac{\chi_n(a_1)\chi_n(a_2)\chi_n(a_3)}{\chi_n(q^{1/2})},\label{sci-trifund-sum}
\end{equation}where \begin{equation}
K(a)^{-1}=\prod_{n\ge 0} \left[
(1-q^{n+1}) \prod_{\pm}(1-q^{n+1} a^{\pm2})
\right]
\end{equation} is the same function introduced in \eqref{sci-K}, \begin{equation}
K_0{}^{-1}= \prod_{n\ge 0} (1-q^{2+n}),
\end{equation} and \begin{equation}
\chi_n(a)=a^{n-1}+a^{n-3} + \cdots + a^{3-n} + a^{1-n}
\end{equation} is the character of $(a,1/a)\in \TASU(2)$ in the $n$-dimensional irreducible representation of $\TASU(2)$.  At this point,  the reader should try  to prove this equality \eqref{sci-trifund-sum}, or at least check it by expanding both sides to, say,  $O(q^3)$.

We see that this infinite sum expression is equal to the amplitude of the three-holed sphere of the 2d $q$-deformed $\TASU(2)$ Yang-Mills, introduced in Sec.~\ref{2dqdeformedfinal}, in the zero area limit, apart from the prefactor involving $K(a_i)$  and $K(0)$. 

Next, consider the four-punctured sphere. The corresponding four-dimensional theory is obtained by taking two half-hypermultiplets of $\TASU(2)_1\times \TASU(2)_2\times \TASU(2)$ and
$\TASU(2)\times \TASU(2)_3\times \TASU(2)_4$,
and coupling them by a dynamical $\TASU(2)$ gauge multiplet. Correspondingly, the partition function on $S^1\times S^3$ is given by the formula \eqref{sci-gauge}:\begin{align}
Z(a_1,a_2;a_3,a_4)&=\frac12\oint \frac{dz}{2\pi iz}(1-z^2)(1-\frac1{z^2})
K(z)^{-2}Z(a_1,a_2,z)Z(z,a_3,a_4)\\
&= \frac12\oint \frac{dz}{2\pi iz}(1-z^2)(1-\frac1{z^2})
K(z)^{-2} \nonumber\\
&\qquad \times
\prod_{\pm\pm\pm}\prod_{n\ge 0} \frac1{1-q^{n+1/2} a_1^{\pm1}a_2^{\pm1}z^{\pm1} }
\prod_{\pm\pm\pm}\prod_{n\ge 0} \frac1{1-q^{n+1/2} z^{\pm1}a_3^{\pm1}a_4^{\pm1} }
\end{align} where we used the infinite product form of the contribution from the half-hypermultiplet, \eqref{sci-trifund-prod}. In this form it is not clear that this expression is symmetric under the permutation of variables $a_{1,2,3,4}$.

Plugging in the infinite-sum form \eqref{sci-trifund-sum} instead, one finds \begin{align}
Z(a_1,a_2;a_3,a_4)
&= \frac12\oint \frac{dz}{2\pi iz}(1-z^2)(1-\frac1{z^2})
\frac{K(a_1)K(a_2)K(a_3)K(a_4)}{K_0{}^2} \nonumber\\
& \qquad \times 
\sum_{n\ge 0} \frac{\chi_n(a_1)\chi_n(a_2)\chi_n(z)}{\chi_n(q^{1/2})} 
\sum_{m\ge 0} \frac{\chi_m(z)\chi_m(a_2)\chi_n(a_3)}{\chi_m(q^{1/2})} \\
&=
\frac{K(a_1)K(a_2)K(a_3)K(a_4)}{K_0{}^2}
\sum_{n\ge 0} \frac{\chi_n(a_1)\chi_n(a_2)\chi_m(a_3)\chi_n(a_4)}{\chi_n(q^{1/2})^2}.\label{sci-four}
\end{align} Here, we used the orthogonality of the $\TASU(2)$ characters under the natural measure \begin{equation}
\frac12\oint\frac{dz}{2\pi iz} (1-z^2)(1-\frac1{z^2}) \chi_n(z) \chi_m(z)=\delta_{mn}
\end{equation} which is just a special version of \eqref{grouportho}.  The factor $(1-z^2)(1-z^{-2})$ is a measure factor introduced by restricting the group variable $z\in \TASU(2)$ into its Cartan torus. 

We find that the partition function \eqref{sci-four} is again equal to the amplitude of the four-holed sphere of the $q$-deformed $\TASU(2)$ Yang-Mills, apart from the prefactor involving $K(a)$ and $K_0$, and the fact that the area $A$ needs to be set to zero. 

This computation can easily be generalized to arbitrary class S theories of type $\TASU(2)$. 
On the four-dimensional side, we consider the partition function on $S^1\times S^3$ of the theory associated to a Riemann surface of genus $g$ with $n$ punctures, where we put the holonomy $(a_i,1/a_i)\in \TASU(2)_i$ for the flavor symmetry for the $i$-th puncture.
On the two-dimensional side, we take the $q$-deformed $\TASU(2)$ Yang-Mills on the same Riemann surface, where we regard punctures as holes, with the holonomy $(a_i,1/a_i)\in \TASU(2)$ specified around the $i$-th hole. Then we have the general relation \begin{equation}
Z_{4d,g,n}(a_i)  = \frac{\prod_i K(a_i)}{K_0{}^{2g-2+n}} Z_{2d,g,n,A=0}(a_i).
\end{equation}

The factor $K(a)/K_0$ for each puncture can be absorbed into a redefinition of the hole-introducing operator on the two-dimensional side. Similarly, we can always add a local counter-term in a two-dimensional non-gravitational theory given by \begin{equation}
S_{2d} \to S_{2d} + c \int \sqrt{\mathsf{g}} R_{\mathsf{g}}  = S_{2d}+c(2-2g)
\end{equation} where $R_{\mathsf{g}}$ is the curvature scalar of the two-dimensional metric. This can absorb the factor $K_0{}^{2-2g}$.  Therefore, we conclude that the $S^1\times S^3$ partition function of the class S theory of type $\TASU(2)$ associated to a punctured Riemann surface is always given by the partition function of the $q$-deformed $\TASU(2)$ Yang-Mills considered on the same punctured Riemann surface. This is the correspondence first found in \cite{TAGadde:2011ik}.

\subsection{$S^4$ and the Liouville theory}\label{sec:2d4d-liou}
Let us next consider the partition function on $S^4$, which will be a function of the parameters $\epsilon_{1,2}$ in the supergravity background, the masses $m_i$, and the complexified gauge couplings $\tau_i$. Note that $\epsilon_{1,2}$ and $m_i$ all have mass dimension 1. We fix the scale by demanding $\epsilon_1\epsilon_2=1$.

Take a three-punctured sphere. The corresponding four-dimensional theory associated in Sec.~\ref{sec:4} is a half-hypermultiplet in $2\otimes 2\otimes 2$ of $\TASU(2)_1\times \TASU(2)_2\times \TASU(2)_3$. 
Its partition function $Z_{S^4}^\text{trifund}(m_1,m_2,m_3)$ on $S^4$, when the mass parameters of $\TASU(2)_i$ is $(m_i/2,-m_i/2)\in \TASU(2)_i$, is given by $|Z_\text{n.h.}(m_1,m_2,m_3)|^2$ as in \eqref{pestun-integral}. Here 
the contribution from the northern hemisphere, $Z_\text{n.h.}^\text{trifund}(m_1,m_2,m_3)$ is given in \eqref{nek-hyper}:
\begin{equation}
Z_\text{n.h.}^\text{trifund}(m_1,m_2,m_3)=\prod_{\pm\pm}
\Gamma_2(\frac{m_1\pm m_2\pm m_3}2 + \frac{\epsilon_1+\epsilon_2}2|\epsilon_1,\epsilon_2)
\end{equation}  and therefore we have \begin{equation}
Z_{S^4}^\text{trifund}(m_1,m_2,m_3)=
\prod_{\pm\pm\pm}
\Gamma_2(\frac{\pm m_1\pm m_2\pm m_3}2 + \frac{\epsilon_1+\epsilon_2}2|\epsilon_1,\epsilon_2).
\end{equation}  Note that we used our convention that $m_i$ are all purely imaginary.

We see that this is equal to the denominator of the three-point function \eqref{DOZZ} of the Liouville theory, under the identification \begin{equation}
\alpha_i = m_i + \frac{b+1/b}2, \qquad (\epsilon_1,\epsilon_2)=(b,\frac1b),
\end{equation} after substituting the definition \eqref{ups} of the function $\Upsilon$. Most of the factors in the numerator can also be accounted for by identifying the two-dimensional operator \begin{equation}
V_{\alpha_i}:=\Upsilon(2\alpha_i) e^{2\alpha_i \phi}
\end{equation} with the puncture with mass $m_i$ in the class S construction.
Then, most of the mass-dependent terms of the Liouville three-point function is in the partition function of the trifundamental half-hypermultiplet. Let us denote this as an equation: \begin{equation}
Z_{S^4}^\text{trifund}(m_1,m_2,m_3) = \vev{V_{\alpha_1}(\infty) V_{\alpha_2}(1) V_{\alpha_3}(0) }_\text{Liouville}
\end{equation} where the equality is up to a multiplication by functions  independent of $m_i$.

Now, let us consider the four-punctured sphere. The corresponding four-dimensional theory is obtained by taking two trifundamentals, one for $\TASU(2)_1\times \TASU(2)_2\times \TASU(2)$ and another for $\TASU(2)\times \TASU(2)_3\times \TASU(3)_4$, and coupling it to an $\TASU(2)$ vector multiplet. The $S^4$ partition function is,
according to \eqref{pestun-integral}, \begin{multline}
Z_{S^4}(m_1,m_2;m_3,m_4;\tau)
=\frac12\int da  |q|^{2 a^2} Z_{S^4}^\text{trifund}(m_1,m_2,a)
Z_{S^4}^\text{trifund}(a,m_3,m_4)  \\
\times
Z_{S^4}^\text{gauge,one-loop}(a) 
|Z_{n.p.}^\text{instanton}(a,m_i,q)|^2.\label{5.18}
\end{multline}  Here, $m_i$ are the mass parameters for $\TASU(2)_i$, $Z_{n.p.}^\text{instanton}(a,m_i,q)$ is the instanton contribution from the north pole, and $Z_{S^4}^\text{gauge,one-loop}(a)$ is the one-loop contribution from the gauge multiplet. Let us start with the gauge one-loop factor, which is \begin{equation}
Z_{S^4}^\text{gauge,one-loop}(a)  = \prod_{\pm} \frac{1}{\Gamma_2(\pm a|\epsilon_1,\epsilon_2)
\Gamma_2(\pm a +\epsilon_1\epsilon_2|\epsilon_1,\epsilon_2)} = \Upsilon(a)\Upsilon(Q-a). 
\end{equation}
Together with the factors $Z_{S^4}^\text{trifund}(m_1,m_2,a)$ and $Z_{S^4}^\text{trifund}(a,m_3,m_4)$,
this provides the product of three-point functions in the expression of the Liouville four-point function \eqref{liouville-four}. 

This strongly suggests that the rest of the factors should match, namely, 
that the conformal block $F$ in \eqref{liouville-four} and the instanton partition function $Z_{n.p.}^\text{instanton}$ should agree:
\begin{equation}
F_{\alpha_1,\alpha_2,ip +Q/2,\alpha_3,\alpha_4}(q)
\quad \text{``}=\text{''} \quad 
Z_{n.p.}^\text{instanton}(a,m_i,q) \label{conf-vs-inst}
\end{equation} under the identification \begin{equation}
\alpha_i = m_i+\frac Q2, \quad
ip + \frac Q2 = a + \frac Q2.
\end{equation}

The equality \eqref{conf-vs-inst} almost works, in the following sense. As already mentioned in Sec.~\ref{sec:S4}, we can only compute the instanton contribution of the $\TAU(2)$ gauge fields, for which no half-hypermultiplet exists, since the doublet and the anti-doublet are different for $\TAU(2)$. 
The half-hypermultiplet in the trifundamental $\TASU(2)_1\times \TASU(2)_2\times \TASU(2)$ with mass parameters $m_1,m_2$ is regarded as two fundamental hypermultiplets in $\TAU(2)$, with masses $m_1+m_2$ and $-m_1+m_2$.  The contribution from these two fundamental hypermultiplets can be computed using the formula 
\eqref{nek-bifund} for the bifundamental of $\TAU(2)\times \TAU(2)$. 
Similarly, we regard the half-hypermultiplet in the trifundamental $\TASU(2)\times \TASU(2)_3\times \TASU(2)_4$ as two fundamental hypermultiplets in $\TAU(2)$, with masses $m_3+m_4$ and $m_3-m_4$. 

The concrete version of \eqref{conf-vs-inst} is then \begin{equation}
Z_{n.p.,\text{$\TAU(2)$ formulation}}^\text{instanton}(a,m_i,q) 
= (1-q)^{2(m_2-Q/2)(Q/2-m_3)}
F_{\alpha_1,\alpha_2,ip +Q/2,\alpha_3,\alpha_4}(z).
\end{equation} The factor which is a fractional power of $(1-q)$ is the spurious contribution that is already alluded to in Sec.~\ref{sec:S4}. The reader is advised at this point to compute both sides of the equation above to order $q^2$ and check the equality. This can be done by using only the formulas already quoted in this review.\footnote{The author joined the collaboration that led to \cite{TAAlday:2009aq} at a rather late stage, and  his contribution was only  to provide the Mathematica code that does the instanton counting that was used to check this equality.}

%The author thinks that 
This spurious prefactor should be dropped for the physically correct version of the partition function, so we just set \begin{equation}
Z_{n.p.,\text{corrected}}^\text{instanton}(a,m_i,q) 
= 
F_{\alpha_1,\alpha_2,ip +Q/2,\alpha_3,\alpha_4}(z).
\end{equation}
We then have the equality \begin{equation}
Z_{S^4}(m_1,m_2;m_3,m_4;\tau)=\vev{V_{\alpha_1}(\infty) V_{\alpha_2}(1) V_{\alpha_3}(z) V_{\alpha_4}(0)}_\text{Liouville}
\end{equation} where the equality is again up to multiplication by a function independent of $m_i$ and $z$.\footnote{For example, in \eqref{5.18} the integrand involves $|q|^{2a^2}$ while in \eqref{liouville-four} the integrand involves $|q|^{2a^2+Q^2/2}$. Therefore, there is naively a mismatch (among others) of a factor $|q|^{Q^2/2}$ between the two expressions. But note that \Nequals{2} supersymmetric theories on curved spaces such as $S^4$ have local supersymmetric counterterms \cite{TAGerchkovitz:2014gta,TAGomis:2014woa,TAGomis:2015yaa} that can change the partition function by a multiplication of the form $|f(\tau)|$ where $f$ is some holomorphic function. This ambiguity accounts for the mismatch $|q|^{Q^2/2}$. }
%\emph{???What happened to the prefactor $|q|^{\Delta-m_i}$ etc???}

This analysis can be extended to a linear quiver and a circular quiver with $\TASU(2)^n$ gauge group. The restriction comes from our inability to compute the instanton partition function with a trifundamental half-hypermultiplet when all three $\TASU(2)$s couple to dynamical gauge fields. 

\subsection{Six-dimensional interpretation}\label{sec:6}
Let us summarize what we saw so far. 
In Sec.~\ref{sec:4}, we introduced a class of four-dimensional \Nequals{2} theories associated to a Riemann surface with punctures. With $n$ punctures, the corresponding theory has $\TASU(2)^n$ flavor symmetry.
In Sec.~\ref{sec:2d4d-sci}, we computed the partition function of these theories on $S^1\times S^3$, using the formulas reviewed in Sec.~\ref{sec:part-sci}. We found that it is equivalent to the two-dimensional $q$-deformed $\TASU(2)$ Yang-Mills theory on the same Riemann surface that we recalled in Sec.~\ref{sec:2d-gauge}

In Sec.~\ref{sec:2d4d-liou}, we computed the partition function of these theories on $S^4$, using the formulas reviewed in Sec.~\ref{sec:S4}. We found that it is equivalent to the Liouville theory on the same Riemann surface that we recalled in Sec.~\ref{sec:2d-liou}.
How should we understand these correspondences? We already gave the outline in Sec.~\ref{sec:intro}. 
Here let us see slightly more details. 

\subsubsection{6d \Nequals(2,0) theory and 4d class S theories}

We start from the six-dimensional \Nequals{(2,0)} theory of type $\TASU(2)$, and we put it on a Riemann surface $C$.   Let us discuss a few features of this theory that will be important for us. 
When compactified on $S^1$, it becomes the maximally supersymmetric $\TASU(2)$ Yang-Mills theory in five dimensions. The precise sense in which it `becomes' the five-dimensional theory is hotly debated.  One strange fact is that the instanton number of the five-dimensional theory should be identified with the Kaluza-Klein momentum of the six-dimensional theory on $S^1$.  Practically, it is known that, as far as the quantities protected by the supersymmetry are concerned, we just have to include the supersymmetric instanton contributions in the computation. More details of this point can be found in \volcite{KL}. We should definitely \emph{not} include Kaluza-Klein towers of massive fields in five dimensions, as it would lead to overcounting. 

This six-dimensional theory has codimension-2 operators, such that on each one of them we have an $\TASU(2)$ flavor symmetry.  
The reader would surely be already familiar with operators supported on points. Computing correlation functions of these point-operators is almost the first thing we learn in quantum field theory. 
Operators supported on lines are also quite familiar: in a four-dimensional gauge theory with gauge group $G$,  we can consider the trace of the path-ordered exponential of the gauge field along a line $L$, in any representation $R$ of $G$. This defines the Wilson line on $L$ in the representation $R$.  This introduces an external electrically charged particle whose worldline is $L$. Similarly, we can introduce an external magnetically charged particle on a given worldline: this determines a 't Hooft loop operator. 
The codimension-2 operator of the six-dimensional theory is similar: it extends along four directions in the six-dimensional spacetime. 
This four-dimensional subspace can have its own external $\TASU(2)$ background field, coupled to its flavor symmetry. 

So, given a Riemann surface $C$ endowed with a metric and $n$ chosen points $p_i$ on it,  we consider the spacetime of the form $\BR^{1,3}\times C$ and the \Nequals{(2,0)} theory of type $\TASU(2)$ on it, together with a codimension-2 operator on $\BR^{1,3}\times p_i$ for each $i$.  By making the area of $C$ very small, this defines  a four-dimensional theory with $\TASU(2)^n$ flavor symmetry.
To preserve a number of supersymmetry, we use the $\TASO(5)_R$ symmetry of the \Nequals{(2,0)} theory.
The curvature of $C$ makes the spinor bundle over $C$ non-trivial; but the supercharges of the six-dimensional theory is charged not only under the $\TASO(2)$ metric curvature of $C$ but under the $\TASO(5)_R$ symmetry. 
So, we pick an $\TASO(2)$ subgroup of $\TASO(5)_R$, and put a compensating curvature there, so that some of the supercharge lives in the trivial bundle over $C$. This method is often called the partial topological twisting.

The most naive choice would be to take the subgroup $\TASO(2)_R\times \TASO(3)_R\subset \TASO(5)_R$, and use this $\TASO(2)_R$ for the  partial topological twisting. The resulting theory is \Nequals{2} supersymmetric in four dimensions. This partial topological twisting has an additional feature that the most of the computable four-dimensional physics only depends on the total area and on the complex structure of the two-dimensional surface, and not on the detailed choice of the metric. 

This allows us to perform the following operation: we pick a decomposition of the punctured Riemann surface into three-punctured spheres connected by a thin, long cylinders.  For each of the cylinder, we first perform the reduction around $S^1$. Then, we just have the five-dimensional $\TASU(2)$ super Yang-Mills on an edge, which gives a four-dimensional $\TASU(2)$ gauge multiplet. The complexified coupling of the resulting gauge multiplet can be argued to be given by $\tau$, the complex parameter geometrically associated to the cylinder. 
For each of the three-punctured spheres, we have some four-dimensional \Nequals{2} theory with $\TASU(2)^3$ symmetry: this is our favorite theory, i.e.~the half-hypermultiplet in the trifundamental representation of $\TASU(2)^3$. 

The four-dimensional  theory  $\CalS(C)$ thus obtained is determined intrinsically by the two-dimensional Riemann surface $C$ with punctures.  The Riemann surface, however, has multiple decompositions into three-punctured spheres. Each of such decompositions gives rise to a four-dimensional Lagrangian description. Starting with a genus $g$ surface with $n$ punctures, we always have $3g-3+n$ $\TASU(2)$ gauge multiplets and $2g-2+n$ half-hypermultiplets in the trifundamental, but the $3g-3+n$ coupling constants of  two different decompositions are related in a complicated manner, and the $n$ flavor symmetries of various trifundamentals are permuted in an interesting way.  Most often, an elementary field in one Lagrangian description arises as a monopole or a monopole bound state in another Lagrangian description. 

\begin{figure}[h]
\[
\TAinc{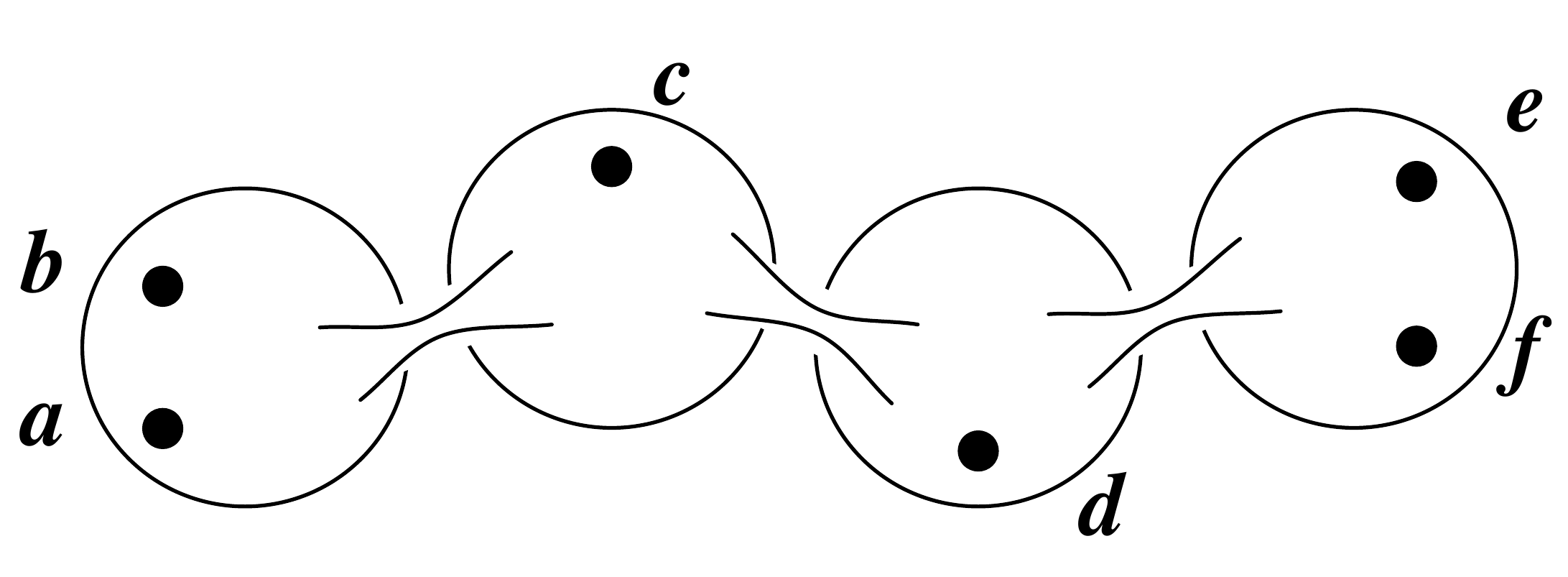}\qquad\TAinc{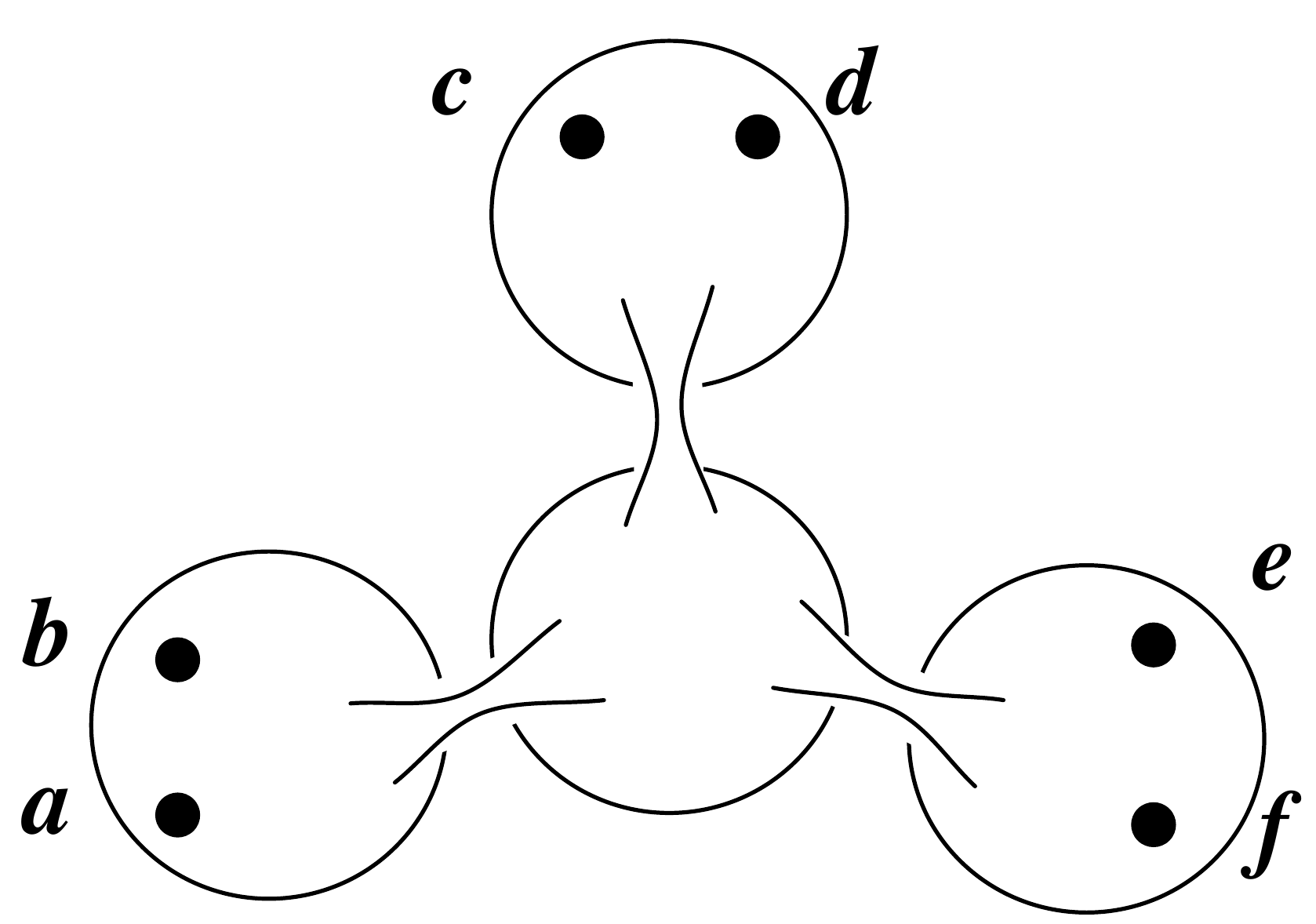}
\]
\caption{A sphere with six punctures under two different decompositions \label{fig:six}}
\end{figure}

As an illustration, consider a sphere with six punctures, see Fig.~\ref{fig:six}. This can be built from four three-punctured spheres connected by three tubes. Therefore the gauge group is $\TASU(2)^3$ and there are four trifundamentals. The decomposition on the left and on the right are rather different, however. For example, one trifundamental coming from the sphere at the center of the decomposition on the right has all three $\TASU(2)$ symmetries coupled to dynamical gauge fields.  But we also know that these two decompositions can be continuously deformed to each other. This represents an S-duality from the 4d Lagrangian perspective.

\subsubsection{Alternative derivations of the correspondences}
\begin{figure}[h]
\[
\TAinc{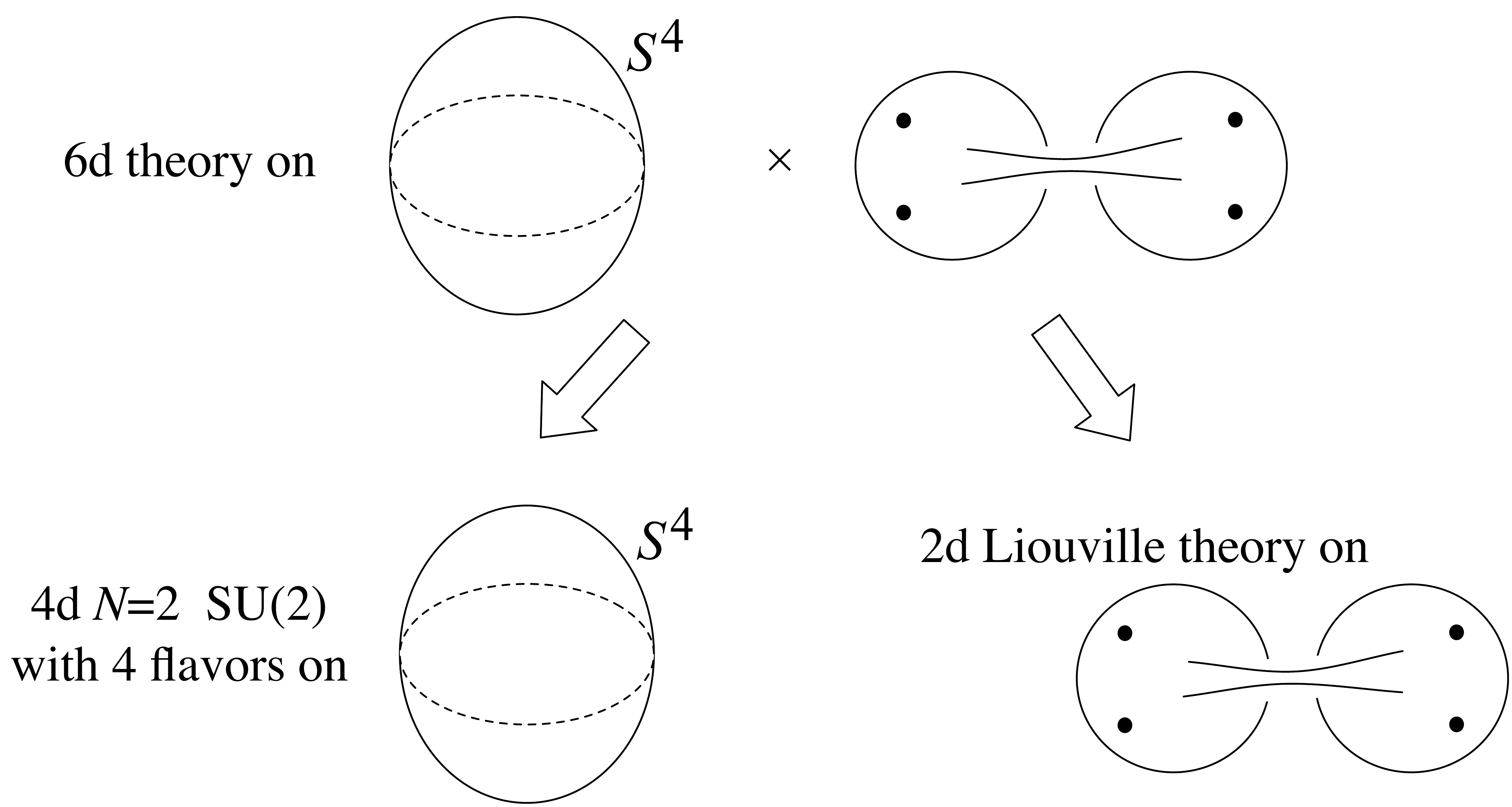}
\]
\caption{6d theory on $S^4\times$ a Riemann surface, and its dimensional reduction \label{fig:relation}}
\end{figure}

In Sec.~\ref{sec:2d4d-sci} and Sec.~\ref{sec:2d4d-liou}, we computed the partition function of $\CalS(C)$ on $S^1\times S^3$ and on $S^4$, and we found that it is given by two-dimensional $q$-deformed $\TASU(2)$ Yang-Mills  on $C$ 
and by the Liouville theory on $C$. In view of the relation $\CalS(X_4)(C_2)=\CalS(C_2)(X_4)$ when the partition function does not depend on the size, we now conclude that \begin{align}
\CalS(S^1\times S^3) &= \text{two-dimensional $q$-deformed $\TASU(2)$ Yang-Mills}, \label{sci}\\
\CalS(S^4) &=\text{Liouville theory}.\label{liou}
\end{align} See Fig.~\ref{fig:relation} for an illustration. 

These two equalities are remarkable: usually in the physics literature, the equality is between numbers. Here, it is between quantum field theories! This, the author believes, shows our steady progress towards a better understanding of quantum field theories in general.

We have arrived at the two equations \eqref{sci}, \eqref{liou} on $\CalS(X_4)$ in a rather roundabout way,  first by studying the four-dimensional theories $\CalS(C)$, second by  computing their partition functions via localization, and finally by identifying the results with the known two-dimensional theories. Is there a more direct way to obtain these relations? 

An obstacle is that we do not know the Lagrangian of the six-dimensional theory. A practical method to proceed is to use an $S^1$ reduction first. 
When $X_4=S^1\times S^3$, we obviously have an $S^1$. We first compactify the six-dimensional theory on this $S^1$. Then we just have the five-dimensional super Yang-Mills on $S^3$. Its path integral can be localized to constant modes along $S^3$, and it essentially gives $\TASU(2)$ Yang-Mills theory on the remaining two-directions. The one-loop fluctuations on $S^3$ then gives a required modification to make it to the $q$-deformed Yang-Mills. This analysis was done in \cite{TAFukuda:2012jr}.

When $X_4=S^4$, we do not directly see an $S^1$. But $S^4$ can be seen as an $S^3$ fibration over a segment, such that at the two ends $S^3$ further degenerates. Now, $S^3$ is an $S^1$ fibration over $S^2$. So we can first reduce the system on this $S^1$.  The system is now the five-dimensional $\TASU(2)$ super Yang-Mills on an $S^2$ fibration over a segment, with a funny boundary condition at both ends. The five-dimensional $\TASU(2)$ super Yang-Mills on $S^2$ can be localized to give an $\TASL(2)$ Chern-Simons on the remaining three dimensions, and the boundary conditions at the end of the segments are such that its compactification down to two dimensions gives rise to the Liouville theory. The details can be found in the paper   \cite{TACordovaJafferis}.

\section{Future directions}\label{sec:conclusions}
At this point we have gone through the very basics of the 2d/4d correspondence: we started from the six-dimensional \Nequals{(2,0)} theory of type $\TASU(2)$.  Its compactification on a punctured Riemann surface can be given a Lagrangian description in terms of $\TASU(2)$ vector multiplets and trifundamental half-hypermultiplets. We can then compute its partition function on $S^1\times S^3$ or on $S^4$. We saw that the partition function equals that of the two-dimensional $q$-deformed $\TASU(2)$ Yang-Mills or that of the Liouville theory, respectively.  We now have a way to understand this result by directly studying the six-dimensional theory on $S^1\times S^3$ or on $S^4$. 

There are still gaps in our understanding in this basic case; and there are many avenues of generalizations.  Let us conclude this review by going over these points. It would be a great pleasure for some of the readers to get involved and solve some of the problems mentioned; it would be even more fantastic if some would open up new directions not even mentioned here.

\subsection{Two unsolved problems in the $\TASU(2)$ case}
In this most basic case, most of the mathematical relations, that are just claimed to hold in this review, have been even rigorously proved. The one big gap in our understanding on the Liouville theory side is in the localization of general class S theory of type $\TASU(2)$ on $S^4$: we still do not know how to obtain the instanton contribution from a genuine half-hypermultiplet in the trifundamental of $\TASU(2)^3$, see e.g.~ \cite{TAHollands:2011zc}. Due to this problem, the $S^4$ partition function can only be computed for linear or  circular quivers.  The author hopes an interested reader would find a way to proceed on this point.

On the side of $S^1\times S^3$ partition function, we only discussed a one-parameter slice of the general superconformal indices with three parameters $(p,q,t)$.  
In the two-parameter slice $(p=0,q,t)$, it is known that the $S^1\times S^3$ partition function gives the $(q,t)$-deformed Yang-Mills theory, where the characters $\chi_\lambda(a)$ in the Yang-Mills partition function are replaced by the Macdonald polynomials $P_\lambda(a|q,t)$.  In the most general three-parameter case, the basic associativity of the four-point function was shown mathematically, but not as we did by rewriting the infinite product into an infinite sum, over a further generalization $P_\lambda(a|p,q,t)$ of Macdonald polynomials, see e.g.~\cite{TARazamat:2013qfa}.

\subsection{More objects in the $\TASU(2)$ case}
So far in this review, we only used one particular type of codimension-2 operators of the six-dimensional \Nequals{(2,0)} theory of type $\TASU(2)$.  It preserves the superconformal symmetry that maps the worldvolume of the operator to itself; such operator is often called regular.  Compactifications on a Riemann surface with regular punctures give superconformal theories in four dimensions, and we discussed them exclusively up to this point. 

There are also irregular codimension-2 operators, that breaks the superconformal symmetry. Compactifications with irregular punctures can lead to asymptotically-free theories in four dimensions, and also superconformal theories of a rather different type, called Argyres-Douglas theories.  The $S^4$ partition function of asymptotically-free theories can be written down, but that of the Argyres-Douglas theories are not well understood from the 4d point of view.  The corresponding Liouville correlators are being  explored, see e.g.~\cite{TAGaiotto:2012sf,TAGomis:2014eya}.  There was a progress in the understanding of the $S^1\times S^3$ partition functions of the Argyres-Douglas theories in 2015, see e.g.~\cite{TABuican:2015ina,TACordova:2015nma}.

The six-dimensional \Nequals{(2,0)} theory of type $\TASU(2)$ also has codimension-4 operators, labeled by irreducible $\TASU(2)$ representations. The ones labeled by trivial 1-dimensional representation are trivial, and the basic nontrivial ones are labeled by the doublet representation.  Recall the basic set-up of the 2d/4d correspondence, where we have the six-dimensional theory on $X_4\times C_2$. As codimension-4 operators have two-dimensional world-volume, we can have the following three situations, roughly speaking: \begin{itemize}
\item The operator extends in two directions in $X_4$,  and sits at a point in $C_2$. This gives a so-called surface operator of the four-dimensional theory. Its gauge theory description is by now well understood. In the Liouville theory this is an insertion of a degenerate operator. Its manifestation in the $q$-deformed Yang-Mills was studied in \cite{TAAlday:2013kda}.
\item The operator extends in one direction in $X_4$,  and in one direction in $C_2$.  This gives a line operator of the four-dimensional theory. In fact, all possible line operators of the class S theories of type $\TASU(2)$ can be nicely described in this way.  They also give line operators in  the Liouville theory \cite{TAAlday:2009fs,TADrukker:2009id}.
\item The operator sits at a point in $X_4$, and covers the entire $C_2$. This just gives a point-operator of the four-dimensional theory. When $C_2$ is the three-punctured sphere, this operator is the trifundamental operator itself.
\end{itemize}
These appear to be more or less understood; they are more interesting when we generalize from $\TASU(2)$ to something larger. 

The codimension-2 operators can also be put into this setup in various ways: \begin{itemize}
\item The operator extends in four directions in $X_4$, and sits at a point in $C_2$. This is the puncture we have been talking about, and changes the four-dimensional theory. 
\item The operator extends in three directions in $X_4$, and wraps a line in $C_2$.  This should give a domain wall operator in the class S theory, but does not appear to have been studied. 
\item The operator extends in two directions in $X_4$, and wraps the entire $C_2$.  This gives a surface operator in the four-dimensional theory, and changes the theory on $C_2$. When $X_4=S^4$, the theory on $C_2$ is the $SL(2)$ WZW theory \cite{TANawataToAppear}; the instanton contribution was studied earlier \cite{TAAlday:2010vg}.
\end{itemize}

\subsection{Using 6d \Nequals{(2,0)} theories of other types}
So far we only discussed the case where we start from the 6d \Nequals{(2,0)} theory of type $\TASU(2)$.
The 6d \Nequals{(2,0)} theories are believed to fall into the ADE classification, i.e.~there are theories of type $G=A_n$, $D_n$ and $E_{n=6,7,8}$.  We can put these theories on $X_4\times C_2$, and have a lot of fun working out the resulting 2d/4d correspondences. 

Without any additional objects in the setup, $\CalS_{G}(S^1\times S^3)$ gives the two-dimensional $q$-deformed Yang-Mills with gauge group $G$, or its $(q,t)$ and $(p,q,t)$ generalizations. $\CalS_{G}(S^4)$ gives a natural generalization of the Liouville theory, known as the two-dimensional Toda theory of type $G$, which has a two-dimensional symmetry called the $W_G$ algebra, that is a generalization of the Virasoro algebra. 

In the \Nequals{(2,0)} theory of type $G$, there are also codimension-2 and codimension-4 operators. 
The codimension-2 operators can be classified into regular ones and irregular ones. The latter are not very well understood, see e.g.~\cite{TAXie:2012hs}. The former are quite well understood: they are labeled by a homomorphism $\mathfrak{su}(2)\to \mathfrak{g}$, see e.g.~\cite{TAChacaltana:2012zy}. The basic one is when the map sends the whole $\mathfrak{su}(2)$  to zero. This is often called the full codimension-2 operator, and is the one that appears when we split a Riemann surface into two.  All the other regular codimension-2 operator can be obtained by giving a vev to the point-operators living on the codimension-2 operator. 

The four-dimensional theories obtained by putting the six-dimensional \Nequals{(2,0)} theory of type $G$ on a Riemann surface with punctures are called class S theories of type $G$. As always, they can be decomposed into the cylinders that basically give rise to $G$ gauge multiplets in four dimensions, and the four-dimensional theory for each of the three-punctured spheres. The theory corresponding to the sphere with three full punctures is called the $T_G$ theory, or $T_N$ theory when $G=\TASU(N)$.  The $T_2$ theory is then the free theory of a half-hypermultiplet in the trifundamental of $\TASU(2)^3$.  The other $T_G$ theories do not admit, however, any useful Lagrangian description.  In a sense the $\TASU(2)$ case was the exception: the six-dimensional \Nequals{(2,0)} theory itself does not have any useful Lagrangian, and therefore we should not expect one in four dimensions in general. 

This presents a big difficulty in the localization computation, since the localization only applies to the Lagrangian part of the theory.  Fortunately, we can still apply the localization to the vector multiplets coming from the cylinders. For example, from this alone, we can conjecture that the instanton partition function of the $G$ gauge theory when $G$ is one of $A_n$, $D_n$ or $E_{n=6,7,8}$ should be controlled by the $W_G$ algebra. This aroused a not quite insignificant interest in the mathematics community, and is now rigorously proved \cite{TASchiffmannVasserot,TAMaulik:2012wi,TABraverman:2014xca}.
However, the $S^4$ partition function of the $T_G$ theory is not yet known.  Under the 2d/4d correspondence, this should map to the three-point function of the Toda theory of type $G$ with general momenta. This is again not known unless $G$ is of type $\TASU(2)$.  Here we see the conservation law of the difficulty at work: something that is difficult on one side of the correspondence is also difficult on the other side.  At the same time, we can say that a breakthrough on either side of the correspondence will have a huge impact on the other.  The author hopes that an interested reader will do make such an epoch-making step.
The $S^1\times S^3$ partition function of the $T_G$ theory has been deduced from the associativity alone, at least in the two-dimensional $(q,t)$ slice. This information has been used to understand the $T_G$ theory better. 

We can put a regular codimension-2 operator specified by a homomorphism $\phi$ on $S^4\times C_2$ such that it wraps the entire $C_2$ and occupy an $S^2\subset S^4$. This is known to modify the $W_G$ symmetry on $C_2$ to a more general W-algebra $W(G,\phi)$ obtained by the quantum Drinfeld-Sokolov reduction. But the two-dimensional theory itself has not been worked out.

Let us now briefly discuss codimension-4 operators of the six-dimensional theory of type $G$. They are labeled by irreducible representations of $G$.  When three representations $R_{1,2,3}$ have an invariant tensor $\varphi: R_1\otimes R_2\otimes R_3 \to \BC$, we can consider a junction of three codimension-4 operator along a one-dimensional locus. We can place such a junction in $X_4\times C_2$ in many ways; analyzing the setup from the point of view of the four-dimensional theory on $X_4$ and from that of the two-dimensional theory on $C_2$ generate various correspondences, some of which have been worked out.

\subsection{Other spacetimes, other theories}
In this review, we only discussed putting the six-dimensional \Nequals{(2,0)} theory on a particular spacetime of the form $X_4\times C_2$, where we perform the partial topological twist along $C_2$ to have \Nequals{2} supersymmetry on $X_4$. 
In fact the choices we actually used in this review were even more restricted. Namely, we only considered just two cases, $X_4=S^1\times S^3$  and $X_4=S^4$. 
Clearly this very short list can be extended. There are a few works on $X_4=S^1\times S^3/\BZ_k$ and on $X_4=S^4/\BZ_k$. For example, the six-dimensional \Nequals{(2,0)} theory of type $\TASU(2)$ on $S^4/\BZ_2$ is known to lead to the \Nequals{(1,1)} super Liouville theory on the two-dimensional side. 

Another choice is to take compact complex toric surfaces as $X_4$. For this, the final formula of the localization computation was announced by Nekrasov in \cite{TANekrasovLisbon}; the details were recently provided by different authors in  \cite{TABershtein:2015xfa}. The corresponding two-dimensional theories labeled by $X_4$ were not understood yet, though.

We can also consider non-compact complex surfaces as $X_4$. In this case we do not obtain a full-fledged 2d theory; rather, the supersymmetric localization on it gives rise to 2d chiral algebras. The case $X_4=\mathbb{R}^4$ was originally studied by \cite{TANekrasov:2002qd} and gives the instanton contribution we discussed in Sec.~\ref{sec:S4}. There are also various studies when $X_4$ is an ALE space, which was pioneered by \cite{TAFucito:2004ry}.  For a sample of references, see e.g.~\cite{TAFucito:2006kn,TAGriguolo:2006kp,TADijkgraaf:2007fe,TABelavin:2011pp,TABonelli:2011jx,TABelavin:2011tb,TABonelli:2011kv,TAWyllard:2011mn,TAIto:2011mw,TAAlfimov:2011ju,TABelavin:2011sw,TABonelli:2012ny,TABelavin:2012eg,TAIto:2013kpa,TAAlfimov:2013cqa,TAItoyama:2013mca,TABruzzo:2013daa,TAPedrini:2014yoa,TASpodyneiko:2014qsa,TABruzzo:2014jza}.\footnote{The author thanks F. Sala for the help in compiling this list of references.}

We can of course consider other decompositions of the six-dimensional spacetime. For example, we can consider putting the six-dimensional \Nequals{(2,0)} theory on spacetimes of the form $X_{6-d}\times C_d$  where we perform the partial topological twist along $C_{d}$, so that the preserved supersymmetry squared generates an isometry in $X_{6-d}$.  
Note that this setup is not symmetric under the exchange $d\leftrightarrow 6-d$. 
In this review we only discussed the case $d=2$; the other cases $d=3,4$ have been analyzed, mainly for six-dimensional \Nequals{(2,0)} theory of type $\TASU(N)$.  
For example, the reader can find some discussion on the 3d-3d case in \volcite{DI}.
We can have even more fun by including various codimension-2 and codimension-4 operators into this general setup. 

Just considering six-dimensional \Nequals{(2,0)} theories on product manifolds give us such plethora of correspondences of lower-dimensional quantum field theories that are being worked out.  Then  a natural direction would be to look for theories other than six-dimensional \Nequals{(2,0)} theories as the starting point. 
In six dimensions, there are many other \Nequals{(1,0)} supersymmetric theories that are believed to be ultraviolet complete. They are even less understood than \Nequals{(2,0)} theories, but there are recent activities to explore their properties systematically. Once basic features are understood, it should not be impossible to consider them on product manifolds, which hopefully would lead to completely new types of 4d-2d and other correspondences. 

Of course we can start from lower-dimensional, more familiar gauge theories on product manifolds of the form $X_{D-d}\times C_d$. For example, \Nequals{4} super Yang-Mills theory with gauge group $G$ was considered on the spacetime of the form $X_2\times C_2$, with a partial topological twisting along $C_2$, and this  is the basis of Witten's approach to the geometric Langlands correspondence.
The two-dimensional theory one obtains on $X_2$ is a non-linear sigma model on the moduli space of the $G$-Hitchin system on $C_2$, and does not seem to have a nice cut-and-paste description when $C_2$ is split into two, as was possible  in case of the four-dimensional class S theory.  It sounds strange  that a four-dimensional gauge theory with known Lagrangian behaves in a more complicated way upon compactification on $C_2$ than a mysterious six-dimensional theory without Lagrangian behaves. 
The main difference seems to lie in the fact that the possible supersymmetric configuration of the gauge fields on the Riemann surface $C_2$ is quite rich, since $C_2$ can have various nontrivial one-cycles, around which the gauge fields can have nontrivial holonomies.  In the case of the six-dimensional \Nequals{(2,0)} theory, we do not have a useful Lagrangian description; very naively, people say that it is a theory of non-abelian two-form fields. Whatever this statement means, it suggests that there can not be too much supersymmetric configuration on the Riemann surface $C_2$, since there is only one two-cycle in $C_2$.  That said, having an explicit Lagrangian description at the starting point, it should in principle be possible to work out every aspect of the correspondences of the lower-dimensional field theories on $X_{D-d}$ and on $C_d$, which the author thinks worth while to pursue. 

\section*{Acknowledgment}
It is a pleasure to thank Tatsuma Nishioka for carefully reading the draft and giving various useful comments.
This work  is supported  in part by JSPS Grant-in-Aid for Scientific Research No. 25870159 and in part by WPI Initiative, MEXT, Japan at IPMU, the University of Tokyo.

\documentfinish
% \ifx\ifLONG\undefined
% \bibliographystyle{ytphys} 
% \bibliography{TA,review} 
% \end{document} 

% \else
% \input{Tachikawa/TA.bbl}
% \fi